\documentclass[letterpaper]{jpconf}
\usepackage{graphicx}
\begin{document}
\title{Hadronic tau decays and the strong coupling}

\author{Diogo Boito,$^a$ Maarten Golterman,\footnote{Speaker at conference}$^b$ Kim Maltman$^{c,d}$ and\\ Santiago Peris$^e$}
\address{$^a$Instituto de F{\'i}sica de S{\~a}o Carlos, Universidade de S{\~a}o Paulo,
CP 369\\ 13570-970, S{\~a}o Carlos, SP, Brazil}
\address{$^b$Department of Physics and Astronomy, San Francisco State University\\
San Francisco, CA 94132, USA}
\address{$^c$Department of Mathematics and Statistics,
York University\\ Toronto, ON Canada M3J~1P3}
\address{$^d$CSSM, University of Adelaide, Adelaide, SA~5005 Australia}
\address{$^e$Department of Physics and IFAE-BIST, Universitat Aut\`onoma de Barcelona\\
E-08193 Bellaterra, Barcelona, Spain}
\ead{maarten@sfsu.edu}

\begin{abstract}
We provide an overview of recent determinations of the strong coupling,
$\alpha_s$, from hadronic $\tau$ decays.   We contrast two analysis methods,
the ``truncated-OPE approach'' and the ``DV-model approach,'' highlighting the
assumptions going into each of these.   We argue that a  detailed study based on ALEPH data shows the truncated-OPE approach to be quantitatively 
unreliable, while the DV-model approach passes all tests.   New data for
hadronic $\tau$ decays from Belle and Belle-II could provide more stringent
tests of the DV-model approach, and thus potentially lead to a more precise
value of $\alpha_s$ from hadronic $\tau$ decays.
\end{abstract}

\section{Introduction}
\medskip
Let us begin with an overview of values for the strong coupling,
$\alpha_s(m_\tau^2)$ obtained in the most recent analyses of hadronic
$\tau$ decays.   Taking the averages of ``FOPT'' and ``CIPT'' values,\footnote{FOPT stands for fixed-order perturbation theory, and CIPT for
contour-improved perturbation theory; for a review, see for instance Ref.~\cite{jamin2005}.
For comparison we use these averaged values in this section; however, we do not recommend using such averages in general, as this averaging is without theoretical basis.}
the most recent values obtained from ALEPH data \cite{aleph,davier2014} are
\begin{eqnarray}
\label{alphasvalues}
\alpha_s(m_\tau^2)&=&0.332(12)\ ,\qquad\cite{davier2014}\ ,\\
&=&0.328(12)\ ,\qquad\cite{pich2016}\ ,\nonumber\\
&=&0.301(10)\ ,\qquad\cite{boito2015}\ ,\nonumber\\
&=&0.309(9)\ ,\qquad\ \,\cite{boito2015}\ \mbox{including OPAL data \cite{opal,boito2012b}}\ .\nonumber
\end{eqnarray}
While the values of Refs.~\cite{davier2014} and \cite{pich2016} are in 
good agreement, there is a significant discrepancy with the values from Ref.~\cite{boito2015}.   It is interesting to also quote some values from the PDG \cite{pdg}, adjusted to the $\tau$ mass:
\begin{eqnarray}
\label{pdgvalues}
\alpha_s(m_\tau^2)&=&0.325(15)\ ,\qquad\mbox{from $\tau$ decays}\ ,\\
&=&0.315(9)\ ,\qquad\ \,\mbox{from lattice QCD}\ ,\nonumber\\
&=&0.314(9)\ ,\qquad\ \,\mbox{world average without $\tau$}\ .\nonumber\end{eqnarray}
While these values are nominally in agreement with each other, the value from
$\tau$ decays is on the high side, and this is clearly related to the higher values in Eq.~(\ref{alphasvalues}), which push up the $\tau$-based PDG average. 
Two natural questions arise regarding the average from $\tau$ decays: (i) is the average actually reliable, and (ii) can we do better?   We address these questions in what follows.

As we will see, the differences originate in the way the non-perturbative
``contamination'' afflicting the determination of $\alpha_s$ from $\tau$ decays
is treated in the analysis.   While the value of the $\tau$ mass is large enough
that one may attempt to apply QCD perturbation theory, it is sufficiently low that
such a determination is not free from non-perturbative effects.
Here, we will compare the analysis methods used in Refs.~\cite{davier2014,pich2016,boito2015}, and argue that the different assumptions
underlying these methods lead to the different values reported in Eq.~(\ref{alphasvalues}).   We will argue that the method used in Refs.~\cite{davier2014,pich2016} is unreliable, while the method used in Ref.~\cite{boito2015} appears to be consistent with the data.   However, also Ref.~\cite{boito2015} employs an assumption
in order to deal with the non-perturbative contamination.   While tests based on
ALEPH and OPAL data support the validity of this assumption, more precise
data, as for instance obtained from Belle and Belle-II, would make it possible to subject also the method of Ref.~\cite{boito2015} to more stringent
consistency tests.

\section{Background}
\medskip
The ratio of the decay width for non-strange hadronic $\tau$ decays to the 
width for $\tau\to\nu_\tau e\overline\nu_e$ can be calculated in QCD
perturbation theory, and takes the form\footnote{In FOPT.}
\begin{equation}
\label{Rtau}
R^{ud}_\tau\equiv\frac{\Gamma(\tau\to\mbox{hadrons}_{ud})}{\Gamma(\tau\to\nu_\tau e\overline\nu_e)}=3S_{\rm EW}|V_{ud}|^2\left(1+\frac{\alpha_s}{\pi}+
5.2\left(\frac{\alpha_s}{\pi}\right)^2+\dots\right)\ ,
\end{equation}
where $S_{\rm EW}$ is a short-distance electroweak correction.   The 
perturbative series has been calculated to order $\alpha_s^4$ \cite{baikov2008}.
In principle, this result can be used to determine $\alpha_s$ from the 
vector ($V$) and axial-vector ($A$) non-strange decays, for which
ALEPH and OPAL obtained relatively precise data.   However, at $\alpha_s(m_\tau^2)/\pi\approx 0.1$, the series is rather slowly converging, indicating that
QCD perturbation theory may not be all there is at the $\tau$ mass.   On the
experimental side, what one sees is not $\tau\to\nu_\tau\mbox{jets}$, but
rather (in the $V$ channel) $\tau\to\nu_\tau\rho^*\to\nu_\tau\mbox{pions}$
with $\rho^*=\rho(770)$, $\rho(1450)$, {\it etc.}, calling into question the
relation between observed hadronic $\tau$ decays and the perturbative
regime.

\begin{figure}[t]
\includegraphics[width=14pc]{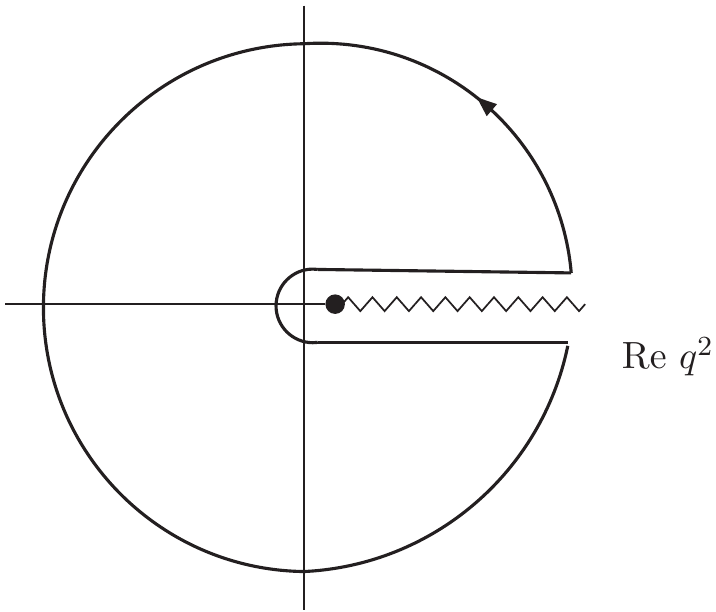}\hspace{2pc}%
\begin{minipage}[b]{14pc}\caption{\label{contour}{\it Contour used in Eq.~(\ref{cauchy}).}}
\end{minipage}
\end{figure}

The finite-energy sum-rule (FESR) approach allows us to look into this
more systematically \cite{braaten1992}.   First, using the optical theorem,
$R^{ud}_\tau$ can be expressed in terms of the $I=1$ vacuum polarization
$\Pi_{V+A}(q^2)$ as \cite{tsai1971}
\begin{equation}
\label{Rtauvacpol}
R^{ud}_\tau=12\pi^2 S_{\rm EW}|V_{ud}|^2\int_0^{m_\tau^2}\frac{ds}{m_\tau^2}
\left(1-\frac{s}{m_\tau^2}\right)^2\left(1+\frac{2s}{m_\tau^2}\right)\rho_{V+A}(s)\ ,
\end{equation}
where $\rho_{V+A}(s)=\frac{1}{\pi}\mbox{Im}\ \Pi_{V+A}(s)$ is the non-strange
inclusive $V+A$ spectral function measured in hadronic $\tau$ decays.  Since
in this talk we will always consider the case $V+A$ we drop the
subscript $V+A$ from now on.   Second, Cauchy's theorem, applied to
the contour shown in Fig.~\ref{contour}, implies that\footnote{The same equation holds for $V$ and $A$ channels separately.}
\begin{equation}
\label{cauchy}
\int_0^{s_0}ds\,w(s)\,\rho(s)=-\frac{1}{2\pi i}\oint_{|z|=s_0}dz\,w(z)\,\Pi(z)\ ,
\end{equation}
where $w(s)$ can be any analytic function.   The idea is now to split $\Pi(z=q^2)$ into a perturbative part $\Pi^{\rm pert}$ and non-perturbative parts,
\begin{equation}
\label{Piz}
\Pi(z)=\Pi^{\rm pert}(z)+\Pi^{\rm OPE}(z)+\Pi^{\rm DV}(z)\ ,
\end{equation}
and substitute this into the contour integral on the right-hand side of
Eq.~(\ref{cauchy}).
Here $\Pi^{\rm OPE}$ is given by the operator product expansion (OPE),\footnote{The coefficients $C_{2k}$ are logarithmically dependent on $q^2$,
but this dependence can be numerically neglected in the application to $\tau$ decays \cite{boito2012}.   For the non-strange channel, $C_2$ is negligibly 
small, being proportional to the squares of the light quark masses.}
\begin{equation}
\label{OPE}
\Pi^{\rm OPE}(q^2)=\frac{C_4}{q^4}-\frac{C_6}{q^6}+\frac{C_8}{q^8}+\dots\ ,
\end{equation}
and $\Pi^{\rm DV}$ represents, by definition, the non-perturbative part not
encapsulated by the OPE.\footnote{In the literature, the perturbative part is often defined to be the leading-order term of the OPE.   Here we will use ``OPE'' to refer to power-correction terms (\ref{OPE}) only.}   Neglecting this latter part, 
and truncating the OPE, for $w(x)$ a polynomial of degree $N$
the right-hand side then depends on a finite number of parameters, $\alpha_s$ 
and $C_{2k}$ for $k=2,\ 3,\ \dots,\ k_{max}=2N+2$.  These parameters can then be fit to the integral
on the left-hand side, computed from the data for the spectral function, as a function of $s_0$.

The physics not represented by the OPE is the physics of resonances, which
are visible in the data along the positive real $z=q^2$ axis in the spectral
function, {\it cf.} Sec.~3 below.   In other words, the OPE does not provide
a reliable representation of the non-perturbative part of $\Pi(z)$ in the region
where the circle with radius $s_0$ crosses the positive real axis in Fig.~\ref{contour}.   As we will see below, this leads to the need to introduce a
non-trivial {\it ansatz} for $\Pi^{\rm DV}(z)$, in order to take into account the
resonance contributions to $\Pi(q^2)$.   Under mild conditions on 
$\rho^{\rm DV}(s)=\frac{1}{\pi}\mbox{Im}\ \Pi^{\rm DV}(s)$, Eq.~(\ref{cauchy})
can be rewritten as \cite{cata2005,cata2008}
\begin{equation}
\label{fesr}
\int_0^{s_0}ds\,w(s)\,\rho(s)
=-\frac{1}{2\pi i}\oint_{|z|=s_0}dz\,w(z)\,\left(\Pi^{\rm pert}(z)+\Pi^{\rm OPE}(z)\right)
-\int_{s_0}^\infty ds\,w(s)\,\rho^{\rm DV}(s)\ .
\end{equation}
If we supply a parametrization for $\rho^{\rm DV}(s)$ in terms of a finite
number of parameters, this constitutes the FESR we will apply to determine
$\alpha_s(m_\tau^2)$ (one of the parameters of the ``theory''
part on the right-hand side) from the experimental data, which allow us to
compute the integrals on the left-hand side.\footnote{The parameters in
$\Pi^{\rm OPE}$ and $\Pi^{\rm DV}$ can be viewed as ``nuisance'' parameters 
in these fits, although they are of physical interest in their own right.}
The representation on the right-hand side is valid for $s_0$ sufficiently large,
and the hope is that it already applies for $s_0$ (not too far) below $m_\tau^2$. 

\section{Data}
\medskip
Plots showing the $V$ and $A$ non-strange spectral functions can be 
found in Fig.~5 of Ref.~\cite{davier2014} and Fig.~5 of Ref.~\cite{opal}.
First, these figures suggest that, short of a complete determination of
the inclusive spectral functions from Belle and Belle-II data, a
determination of the $4\pi$ decay modes might already significantly reduce
the error on the inclusive spectral functions if combined with the ALEPH
data, given the vastly increased statistics of the Belle experiments.   Of
course, a complete determination of the inclusive spectral function from
Belle data is highly desirable as well, given the much higher statistics, and would avoid
potential pitfalls with combining data from different experiments.

\begin{figure}[t]
\includegraphics[width=20pc]{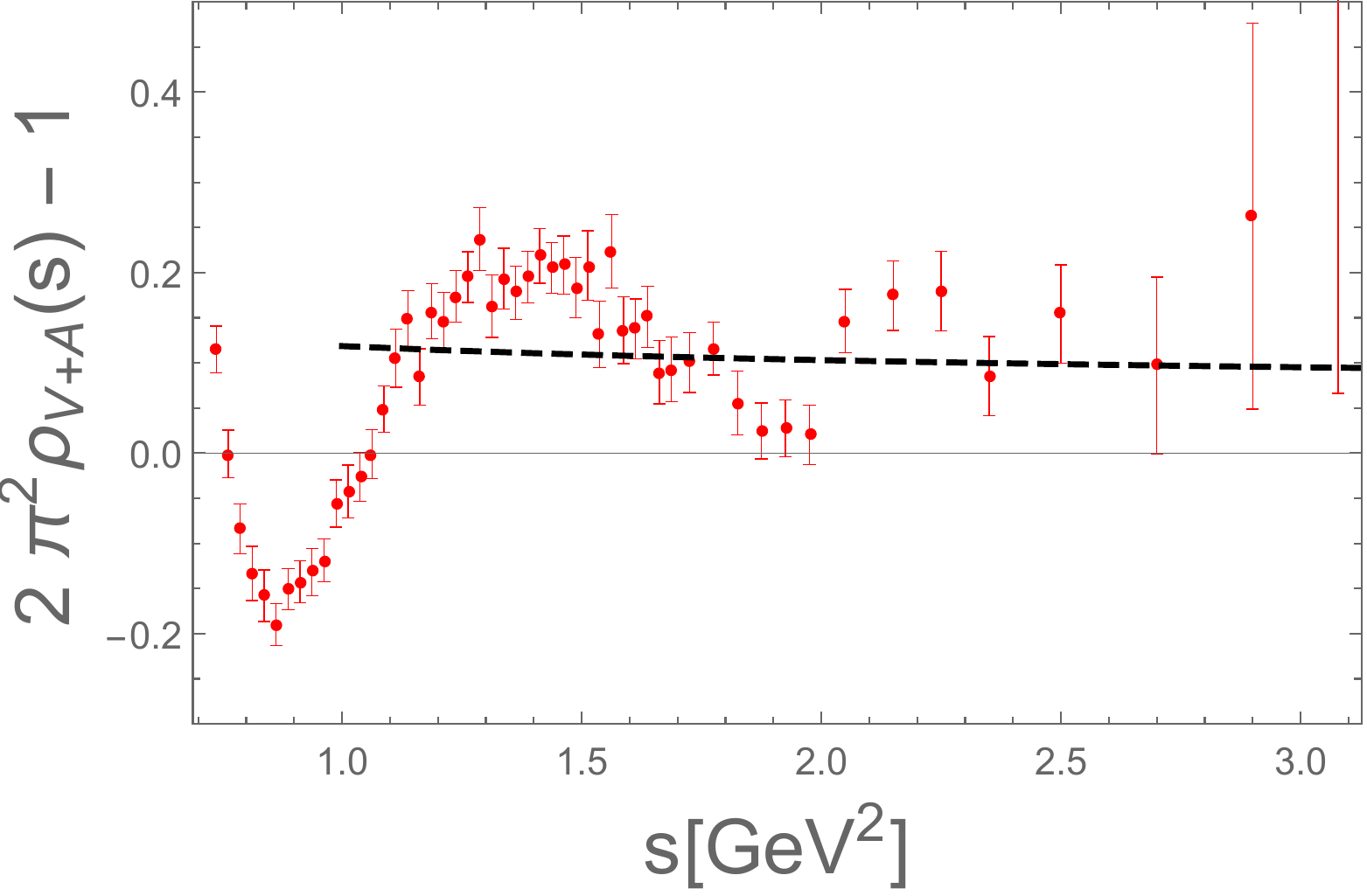}\hspace{2pc}%
\begin{minipage}[b]{16pc}\caption{\label{blowup}{\it Blow-up of the ALEPH data (red data points) in the
large-$s$ region of the $V+A$ non-strange spectral function
($s$ in {\rm GeV}$^2$). 
Black dashed line:  perturbation theory (CIPT) contributions to the spectral function. The $\alpha_s$-independent parton-model contribution to $\rho(s)$ has been subtracted from both.}}
\end{minipage}
\end{figure}

Second, it is interesting to consider the data from the point of view of
what they indicate about non-perturbative effects, in particular, about
resonance effects.   In Fig.~\ref{blowup}, we show the high-$s$ part of the $V+A$ non-strange spectral function measured by ALEPH.
What is shown in this figure is $2\pi^2\rho_{V+A}-1$, {\it i.e.},
the dynamical QCD contribution to the spectral distribution, subtracting the
contribution from QCD without gluons (the parton model).   The
black, dashed curve shows the theoretical representation of the spectral
function without ``duality violations,'' 
obtained from the right-hand side of Eq.~(\ref{fesr}) setting $\rho^{\rm DV}=0$.
The figure shows that the oscillations of the data, due to the presence of
resonances, are not small compared to the contribution from QCD perturbation
theory, and thus need to be taken into account when fitting the data using
Eq.~(\ref{fesr}).

We will turn to the analysis of the data using the two approaches, the
``truncated-OPE'' approach and the ``DV-model'' approach, next.

\section{Truncated-OPE approach}
\medskip
In the truncated-OPE approach \cite{diberder1992}, employed in Refs.~\cite{davier2014,pich2016}, one always takes $s_0=m_\tau^2$, and 
employs the weights ($x=s/s_0$)
\begin{eqnarray}
\label{ALEPHweights}
w_{00}(x)&=&(1-x)^2(1+2x)\ ,\\
w_{10}(x)&=&(1-x)^3(1+2x)\ ,\nonumber\\
w_{11}(x)&=&(1-x)^3(1+2x)x\ ,\nonumber\\
w_{12}(x)&=&(1-x)^3(1+2x)x^2\ ,\nonumber\\
w_{13}(x)&=&(1-x)^3(1+2x)x^3\ .\nonumber
\end{eqnarray}
One assumes that only $C_{4,6,8}$ are non-negligible, and that duality violations, represented
by $\rho^{\rm DV}$ in Eq.~(\ref{fesr}) can be neglected.   We note that the 
weights in Eq.~(\ref{ALEPHweights}) all suppress the region near $s=s_0$
by a double or triple zero, which supports (but does not necessarily validate) the assumption that duality violations can be
neglected.   However, since, from the residue theorem,
\begin{equation}
\label{residue}
\frac{1}{2\pi i}\oint dz\,z^n\,\frac{C_{2k}}{z^k}=C_{2n+2}\,\delta_{k,n+1}\ ,
\end{equation}
we see that, using the weights (\ref{ALEPHweights}), which have values of
$n$ up to $7$, we would need to keep terms in the OPE up to $C_{16}$,
since $2k=2(n+1)=16$ for $n=7$.   The truncated-OPE approach thus 
neglects, without theoretical foundation, the OPE coefficients 
$C_{10,12,14,16}$.   Of course, the practical reason for doing this is that
at $s_0=m_\tau^2$, only 5 data integrals are available, so that a maximum
of 4 parameters, $\alpha_s$, $C_4$, $C_6$ and $C_8$, can be fit.
The truncated-OPE approach thus has two shortcomings:  the truncation of
the OPE, and the neglect of resonance oscillations around perturbation
theory plus the OPE clearly visible in the $V+A$ spectral function, {\it cf.}
Fig.~\ref{blowup}.

\begin{figure}[t]
\begin{minipage}{18pc}
\includegraphics[width=18pc]{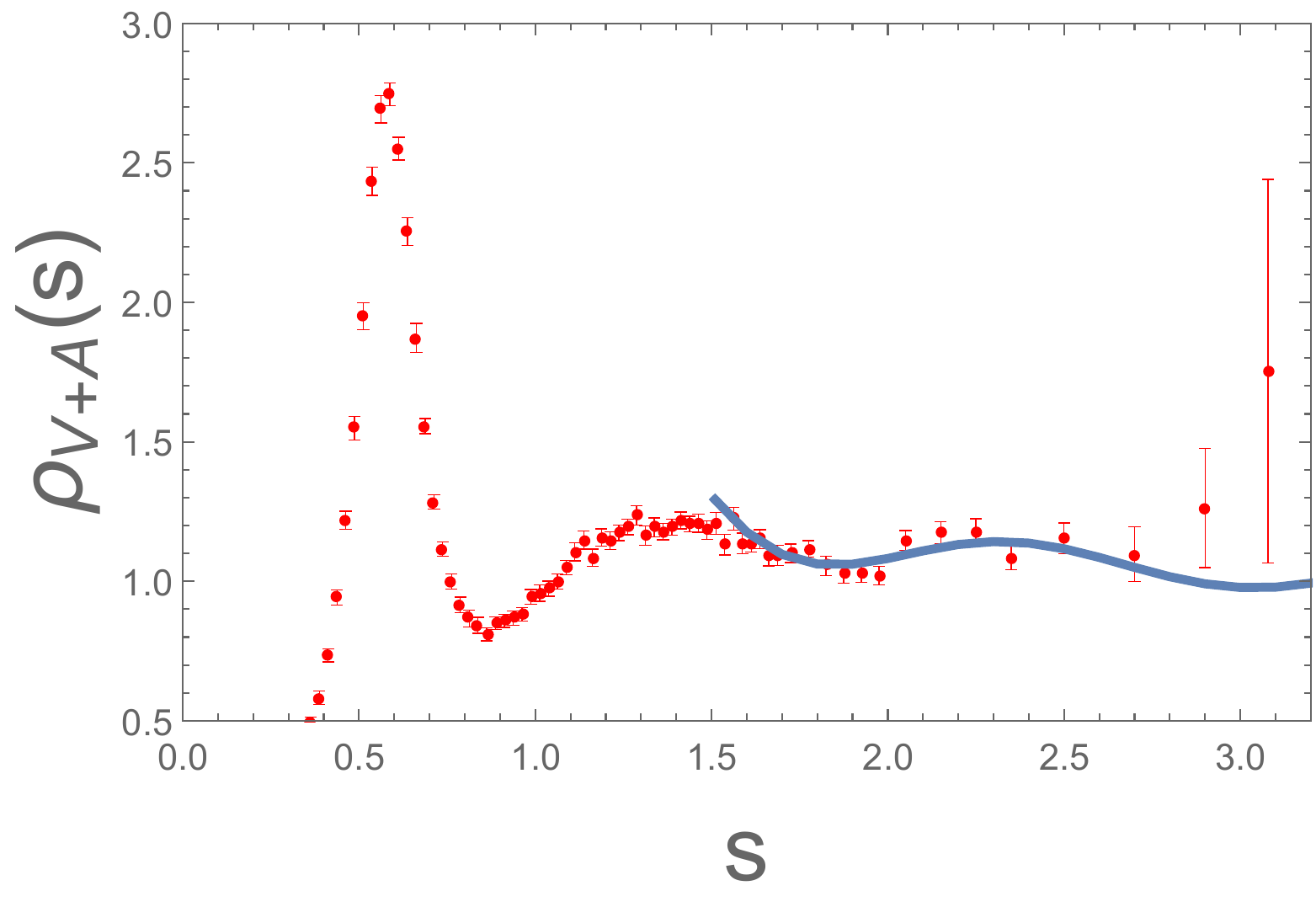}
\end{minipage}\hspace{2pc}%
\begin{minipage}{18pc}
\includegraphics[width=18pc]{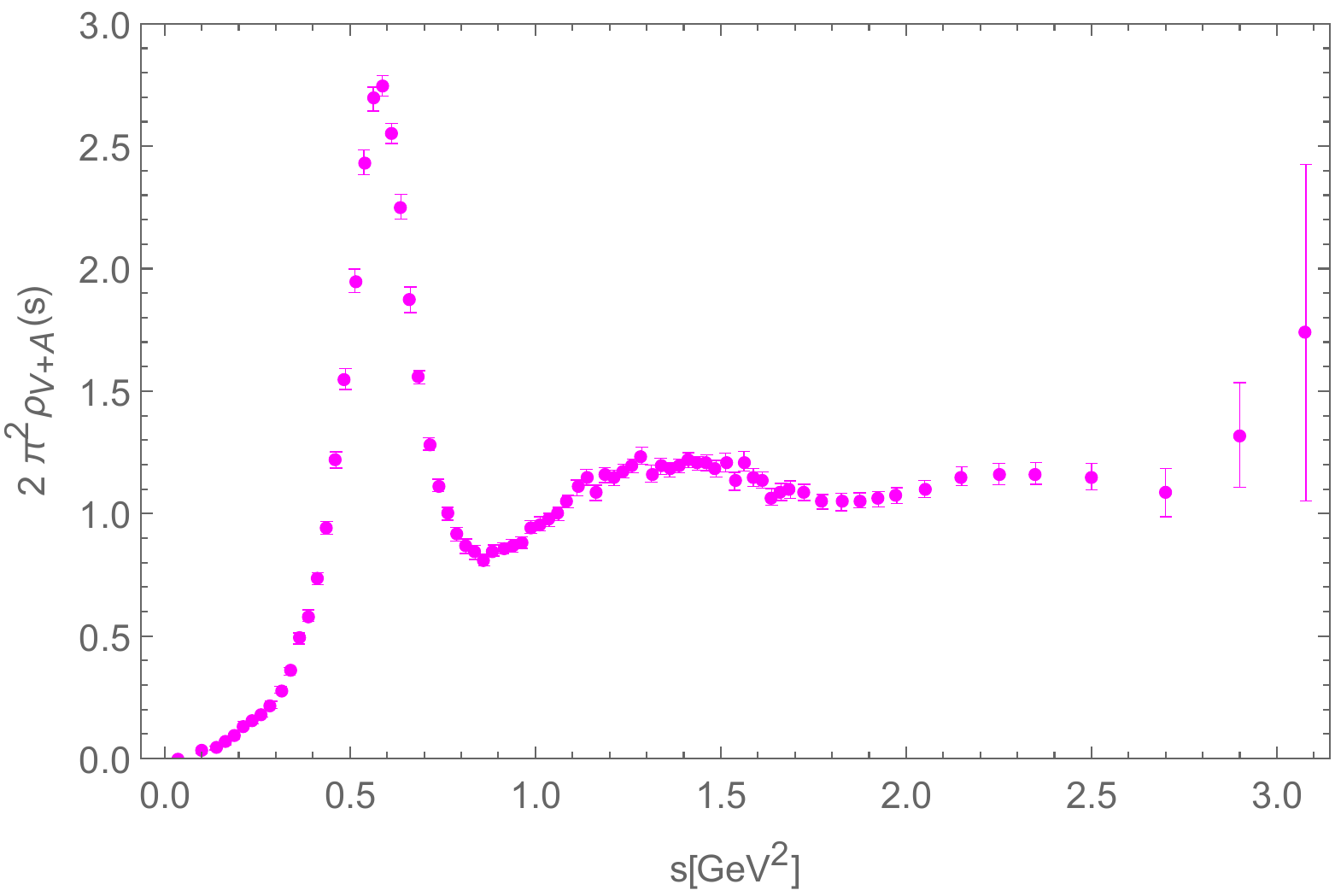}
\end{minipage} 
\caption{\label{fake}{\it Left figure:  true data, with the model curve for $s\ge 1.55$~{\rm GeV}$^2$.   Right figure: fake data for $s\ge 1.55$~{\rm GeV}$^2$,
true data for $s<1.55$~{\it GeV}$^2$.}}
\end{figure}

This approach can be tested, using a set of fake data based on a model with
known values for the parameters \cite{boito2016}.  Using a fit from Ref.~\cite{boito2015}, which describes the $V$, $A$ and thus $V+A$ spectral
functions very well, one can use this fit as a model to generate fake data
from a gaussian distribution with the model as central values, and the 
real-world, ALEPH covariance matrix.   Since the model is only valid for
$s\ge 1.55$~GeV$^2$, we generate fake data only in this region, and use
the true data for $s<1.55$~GeV$^2$.   The model has $\alpha_s(m_\tau^2)=0.312$ (CIPT), and known OPE coefficients (for details, we refer to
Ref.~\cite{boito2016}).

\begin{table}[h!]
\hspace{0cm}\begin{tabular}{|c|c|c|c|c|c|}
\hline
ALEPH &
 $\alpha_s(m_\tau^2)$ & $C_{4}$ (GeV$^4$) & $C_{6}$ (GeV$^6$)& $C_{8}$ (GeV$^8$)& $\chi^2/$dof \\
\hline
true values & 0.312 & 0.0027 & -0.013 & 0.035 & \\
\hline
fake data fit & 0.334(4) & -0.0024(4)  & 0.0007(3) & -0.0008(4) & 0.95/1 \\
\hline
\end{tabular}
    \caption{\label{table1}{\it CIPT fits employing the truncated-OPE approach 
on the fake data, based on the weights of Eq.~(\ref{ALEPHweights}).
By assumption, $C_{10}=C_{12}=C_{14}=C_{16}=0$.
Errors are statistical only.}}
\end{table}%

In Table~\ref{table1} we show the results of fitting the fake data with the
truncated-OPE approach.   We see that the truncated-OPE approach gets
it wrong, and leads to a systematic overestimate of the value of $\alpha_s(m_\tau^2)$, despite an excellent value for the $\chi^2$ per degree of freedom.
In addition, we also note that there is a large difference between the fitted
and true values of the OPE coefficients:  clearly, the fit is trying to compensate
for the ``missing'' contributions from the higher-dimension OPE coefficients 
and/or duality violations that have, by fiat, been set equal to zero.

Reference~\cite{pich2016} considered many other choices of sets of 
weights, in addition to the ``ALEPH'' weights shown in Eq.~(\ref{ALEPHweights}).   In particular,
another set of ``optimal'' weights was considered,
given by
\begin{eqnarray}
\label{optimal}
w^{(2,1)}(x)&=&1-3x^2+2x^3\ ,\\
w^{(2,2)}(x)&=&1-4x^3+3x^4\ ,\nonumber\\
w^{(2,3)}(x)&=&1-5x^4+4x^5\ ,\nonumber\\
w^{(2,4)}(x)&=&1-6x^5+5x^6\ ,\nonumber\\
w^{(2,5)}(x)&=&1-7x^6+6x^7\ .\nonumber
\end{eqnarray}
These weights differ from the ALEPH weights of Eq.~(\ref{ALEPHweights})
in that no linear term in $x$ appears in these weights.  Therefore, they do 
not involve $C_4$, allowing one to fit also $C_{10}$ using the same strategy.
This reduces the reliance on truncation of the OPE by one order, although
the OPE terms proportional to $C_{12}$, $C_{14}$ and $C_{16}$ still need
to be truncated in this approach.  

\begin{table}[h!]
\hspace{0cm}\begin{tabular}{|c|c|c|c|c|c|}
\hline
optimal &
 $\alpha_s(m_\tau^2)$ & $C_{6}$ (GeV$^6$) & $C_{8}$ (GeV$^8$)& $C_{10}$ (GeV$^{10}$)& $\chi^2/$dof \\
\hline
true values & 0.312 & 0.0027 & -0.013 & 0.035 & \\
\hline
fake data fit & 0.334(4) & 0.0008(4)  & -0.0008(5) & 0.0001(3) & 0.92/1 \\
\hline
\end{tabular}
    \caption{\label{table2}{\it CIPT fits employing the truncated-OPE approach 
on the fake data, based on the weights of Eq.~(\ref{optimal}).
By assumption, $C_{12}=C_{14}=C_{16}=0$.
Errors are statistical only.}}
\end{table}%

Table~\ref{table2} shows the result of the fake-data test.  The conclusion is the
same as before:  while the fit results are completely consistent between Tables~\ref{table1} and \ref{table2}, they again disagree with the true model values.
A much more detailed analysis of this problem was given in Ref.~\cite{boito2016}, with the conclusion that the truncated-OPE approach is
inherently unreliable, and thus should no longer be used.

\section{DV-model approach}
\medskip
Our conclusion is that the truncation of the OPE, employed in the approach 
described in the previous section, leads to a systematic bias of the value for
$\alpha_s$ (as well as the OPE coefficients) obtained in that approach.   
Already based on earlier investigations raising doubts about the truncated-OPE
approach \cite{maltman2008,cata2009}, we
employed a simpler choice of weights,\footnote{Note that the sets of weights considered here share $w_{00}(x)=w^{(2,1)}(x)=w_3(x)$.}
\begin{eqnarray}
\label{simple}
w_0(x)&=&1\ ,\\
w_2(x)&=&1-x^2\ ,\nonumber\\
w_3(x)&=&(1-x)^2(1+2x)\ .\nonumber
\end{eqnarray}
It follows from Eq.~(\ref{residue}) that only the coefficients $C_6$ and $C_8$
need to be kept in the OPE.\footnote{This assumes no logarithmic corrections
to the OPE coefficients.   These are suppressed by powers of $\alpha_s$,
and found to be numerically negligible \cite{boito2012}.}  However, the first
weight does not suppress the region $s\approx s_0$ at all, and the second
weight does this only linearly.  It is thus even less likely than in the truncated-OPE
approach that duality violations can be ignored.   We introduce an {\it ansatz} for 
$\rho^{\rm DV}(s)$ in Eq.~(\ref{fesr}); in each of the channels, $V$ and $A$,
we assume that, for large enough $s$, 
\begin{equation}
\label{ansatz}
\rho^{\rm DV}(s)=e^{-\delta-\gamma s}\sin{(\alpha+\beta s)}\ ,
\end{equation}
thus introducing 8 new parameters, $\delta_{V,A}$, $\gamma_{V,A}$,
$\alpha_{V,A}$ and $\beta_{V,A}$ to the fits.   Clearly, we cannot restrict
ourselves to $s_0=m_\tau^2$; instead, we let $s_0$ vary with a minimum
value $s_{min}$, to be determined by the quality of the fits.   It turns out that
the optimal value is $s_{min}\approx 1.55$~GeV$^2$, with good stability
for small variations of this value.

Of course, the use of Eq.~(\ref{ansatz}) implies that this approach is model
dependent.   This is unavoidable, since the physics of resonances is not 
captured by QCD perturbation theory or the OPE, while it is clearly visible 
in the data ({\it cf.} Fig.~\ref{blowup}).   The ingredients going into the
construction of the model
are generally thought of as approximate properties of QCD.   One starts
with assuming 
Regge behavior of the spectrum for $s>s_{min}$, 
which, in the large-$N_c$ limit implies that the spectrum in each channel
looks like $M^2(n)=M^2(0)+\sigma n$, with $n=0,\ 1,\ 2,\ \dots$ and
$\sigma\sim 1$~GeV$^2$.   Relaxing the large-$N_c$ limit, these poles
pick up an imaginary part leading to a decay width $\Gamma(n)\propto
M(n)/N_c$, providing a more realistic picture of the resonance spectrum
in each channel.   A model which incorporates all these features and
satisfies the analyticity constraints of QCD everywhere in the complex
plane exists \cite{cata2005,cata2008,blok1998,bigi1999,golterman2002}, and leads to the {\it ansatz}~(\ref{ansatz}).    

While this approach, the ``DV-model'' approach, is model dependent, it should be stressed that also the truncated-OPE approach is model dependent in the same sense:  it assumes an even simpler model for duality violations, by 
simply setting $\rho^{\rm DV}(s)=0$.   Since the data clearly show the presence of duality violations, this choice of model is almost certainly worse than the
choice~(\ref{ansatz}).   Pinching, {\it i.e.} the suppression of the region
$s\approx s_0$ by the choice of weights~(\ref{ALEPHweights}) or (\ref{optimal}) may help, but without an explicit
model for duality violations this cannot be checked.   As we have shown,
internal consistency between different sets of weights in the truncated-OPE
approach is not sufficient.

Within this framework, many different fits can be performed.   One can choose
any subset of the weights~(\ref{simple}), fit only the $V$ channel or the combined $V$ and $A$ channels, or just the sum $V+A$.   A large suite of
fits has been considered in Refs.~\cite{boito2015,boito2012b,boito2012},
applying this DV-model approach to both the OPAL and ALEPH data.   
The results of these fits show excellent internal consistency.

\begin{figure}[t]
\begin{center}
\includegraphics*[width=7cm]{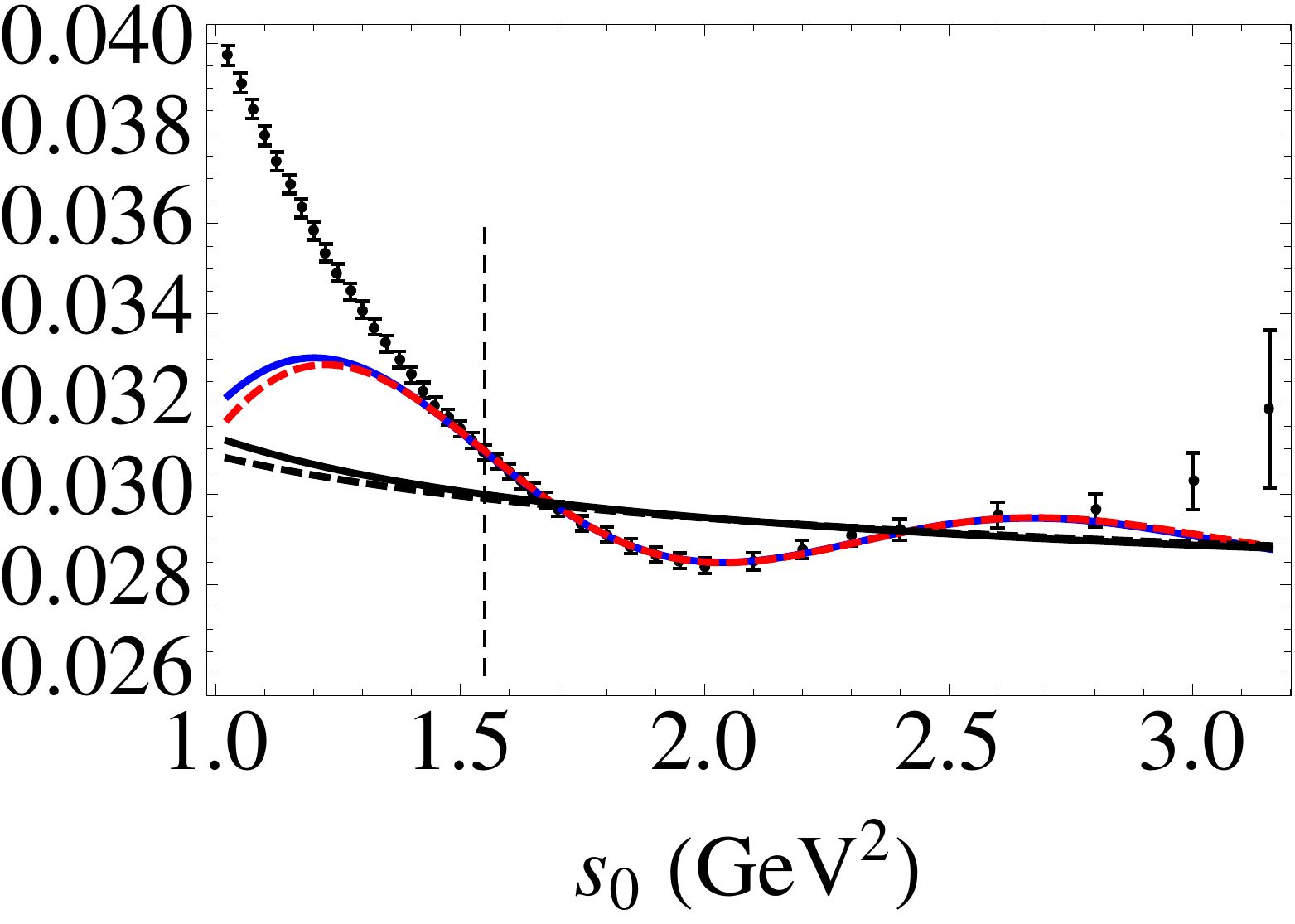}
\hspace{0.5cm}
\includegraphics*[width=7cm]{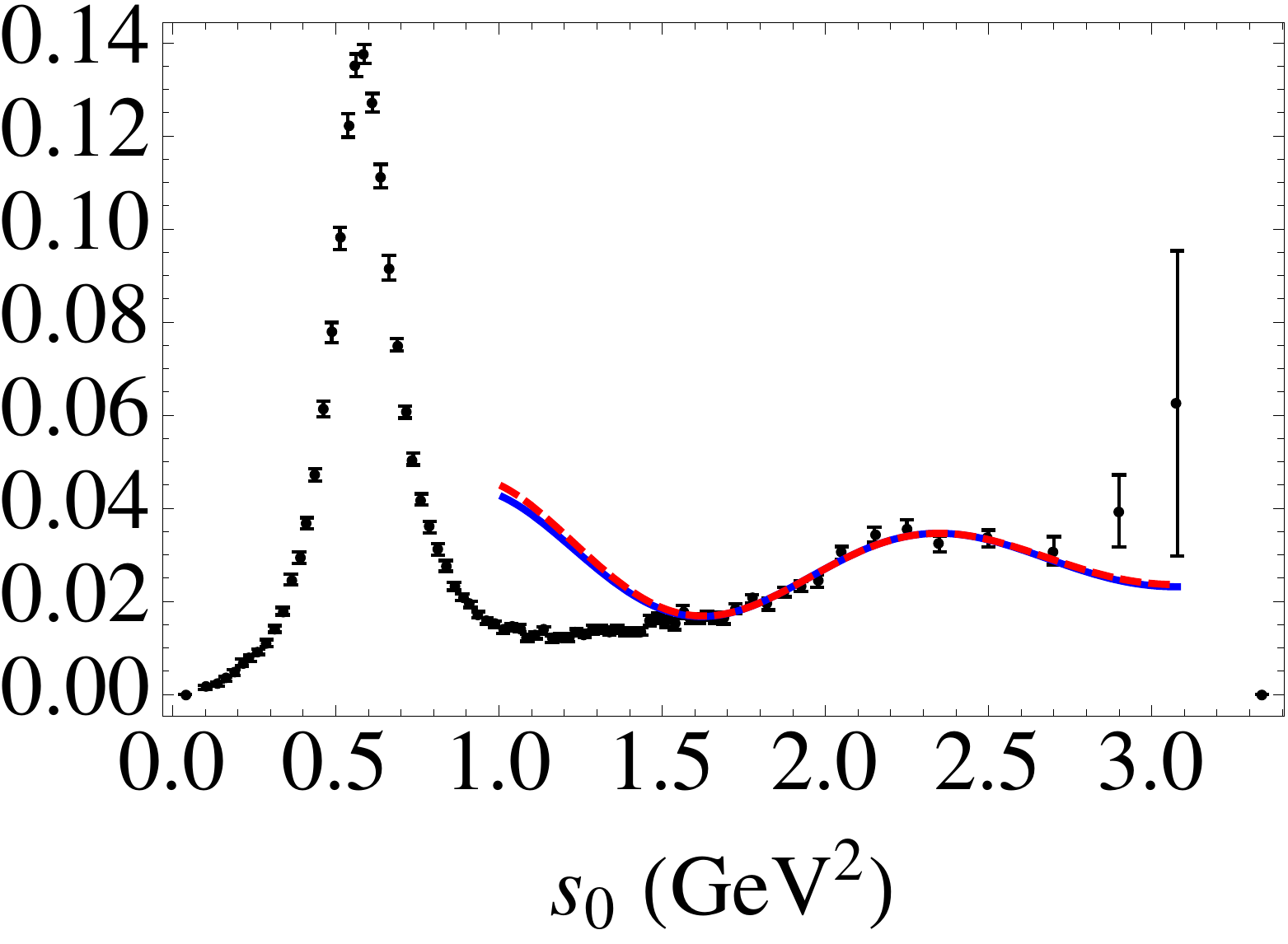}
\end{center}
\begin{quotation}
\caption{\label{CIFOw0fit} {{\it Left panel:
comparison of  $V$-channel, $w_0$ fits to the data. Right panel: comparison of the theoretical
spectral function resulting from this fit with the experimental results.
CIPT fits are shown in red (dashed) and FOPT in blue (solid).
The (much flatter) black curves in the left panel represent the OPE parts of the
fits, {\it i.e.}, the fit results with the DV parts
removed.
The vertical dashed line indicates the location of $s_{\rm min}$.
From Ref.~\cite{boito2015}.
}}}
\end{quotation}
\vspace*{-4ex}
\end{figure}

\begin{figure}[t!]
\begin{center}
\includegraphics*[width=7cm]{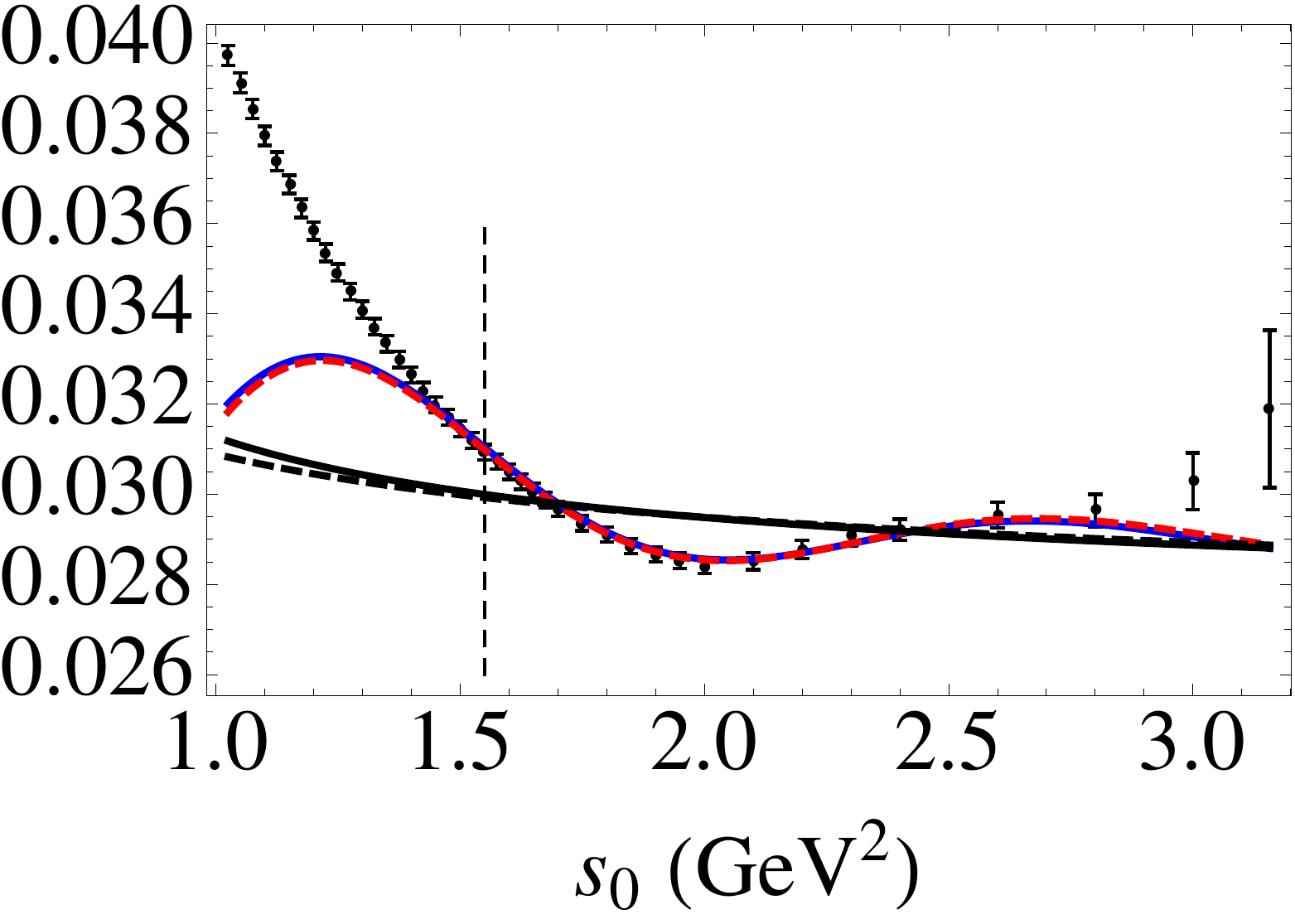}
\hspace{0.5cm}
\includegraphics*[width=7cm]{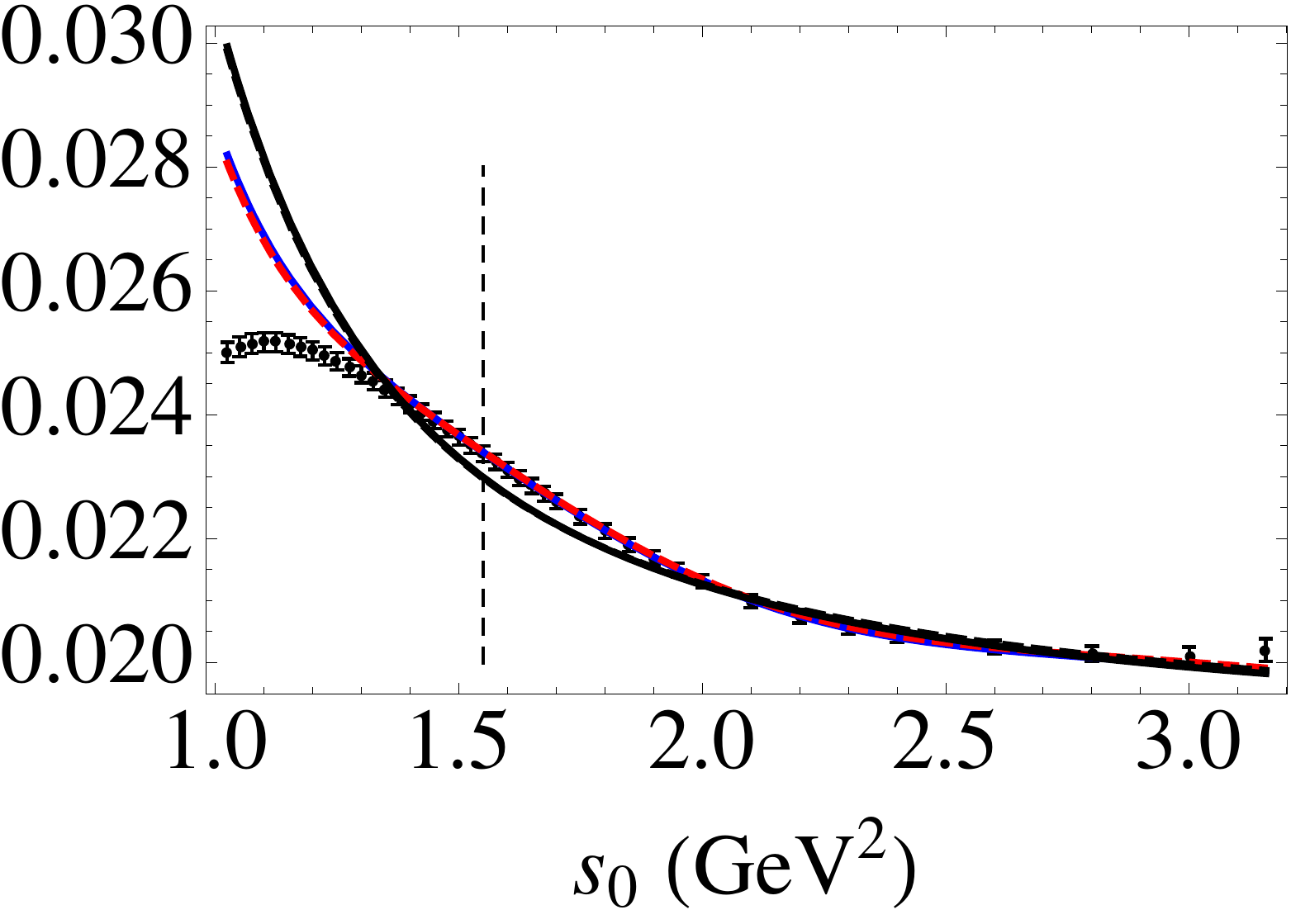}
\vspace{1cm}
\includegraphics*[width=7cm]{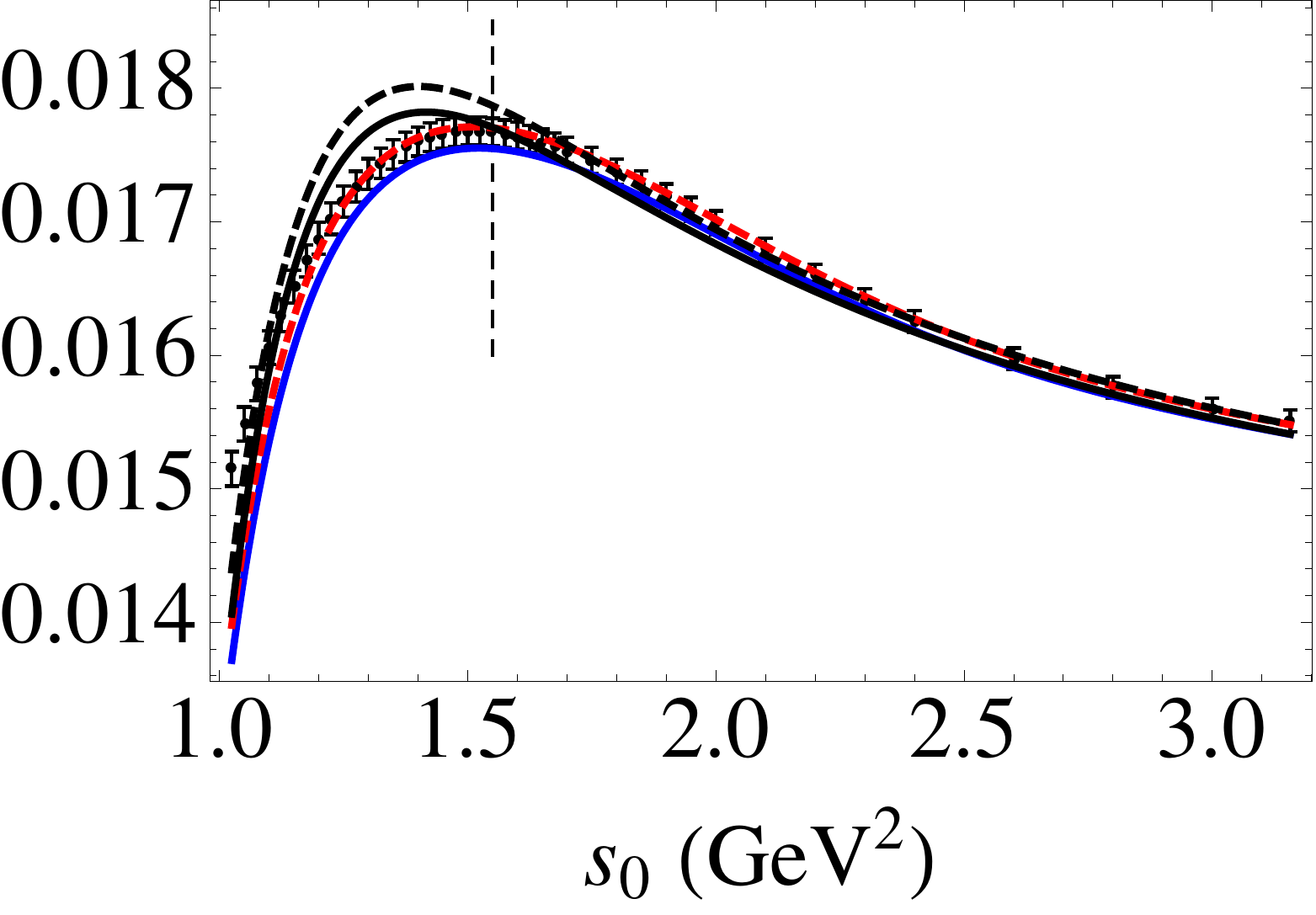}
\hspace{0.5cm}
\includegraphics*[width=7cm]{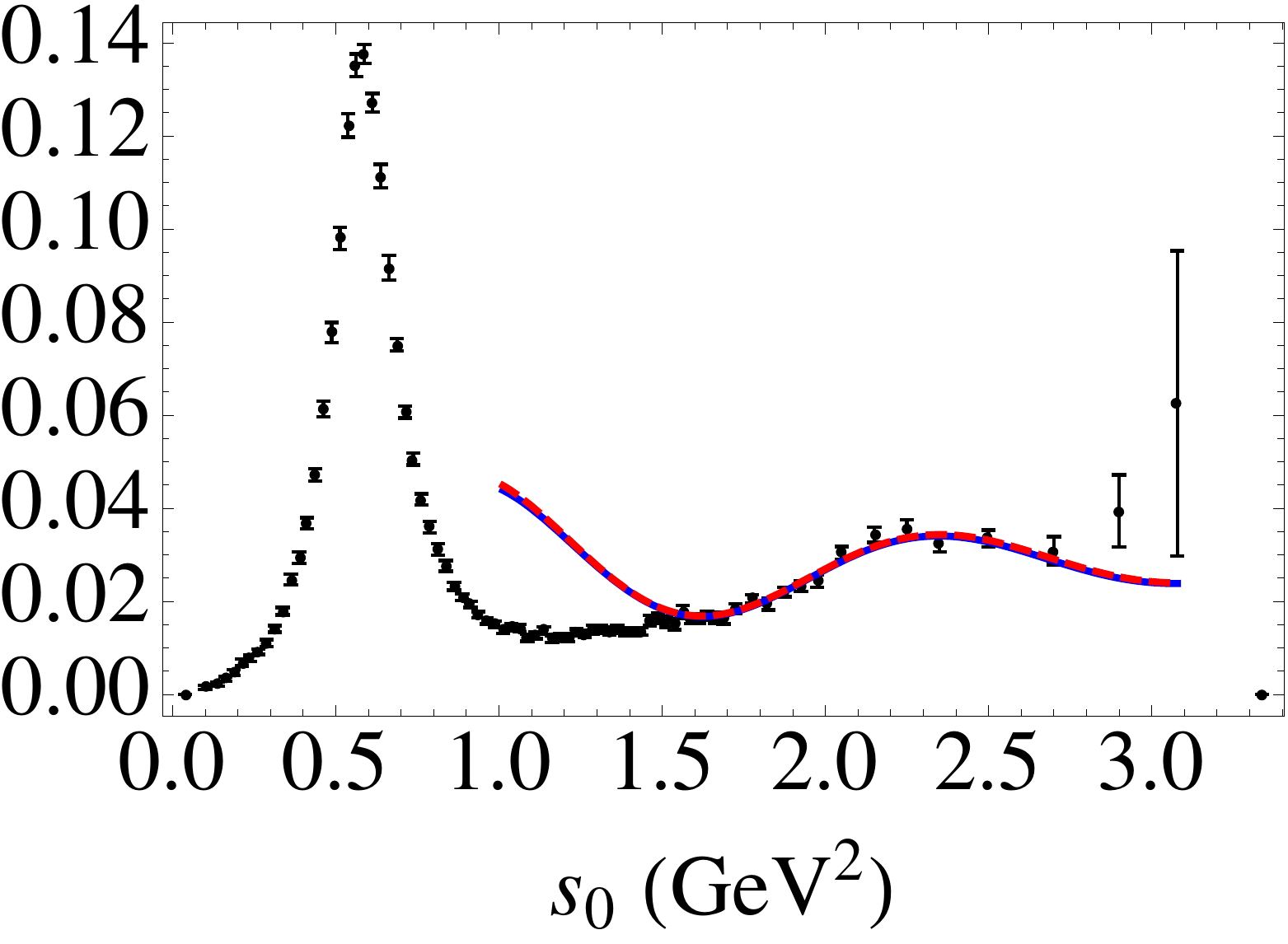}
\end{center}
\vspace*{-6ex}
\begin{quotation}
\caption{\label{CIFOw023fit} {{\it 
Comparison of  $V$ channel fits using the weights $w_0$ (upper left panel),
$w_2$ (upper right panel) and $w_3$ (lower left panel).  CIPT fits are shown in red (dashed) and
FOPT in blue (solid).
Lower right panel: comparison
of the theoretical spectral function resulting
from this fit with the experimental results.
The black curves (which are much
flatter for the $w_0$ case)
represent the OPE parts of the fits. The vertical dashed line
indicates the location of $s_{\rm min}$. From Ref.~\cite{boito2015}.}}}
\end{quotation}
\vspace*{-4ex}
\end{figure}

Figure~\ref{CIFOw0fit} shows an example of the simplest possible fit based on the
DV-model approach, by considering only the $V$ channel, and only the weight
$w_0(x)=1$.   There is excellent agreement between the data and the fits,
shown in the left panel.   The spectral function following from this fit is shown 
in the right panel, along with the ALEPH data.   Another example of a fit 
following this approach, now involving all weights in Eq.~(\ref{simple}), again
considering only the $V$ channel, is shown in Fig.~\ref{CIFOw023fit}.

For many other fits, also involving the $A$ channel, and many consistency checks, we refer to Ref.~\cite{boito2015}.   The results we obtain, from the
ALEPH data, for $\alpha_s$ are
\begin{eqnarray}
\label{results}
\alpha_s(m_\tau^2)&=&0.296(10)\ ,\qquad\mbox{FOPT}\ ,\\
&=&0.310(14)\ ,\qquad\mbox{CIPT}\ .
\end{eqnarray}
The average of these leads to the third value reported in Eq.~(\ref{alphasvalues});
averaging this with the OPAL-based value obtained in Ref.~\cite{boito2012b}
yields the fourth value reported in this equation.

\begin{figure}[t!]
\includegraphics*[width=7.5cm]{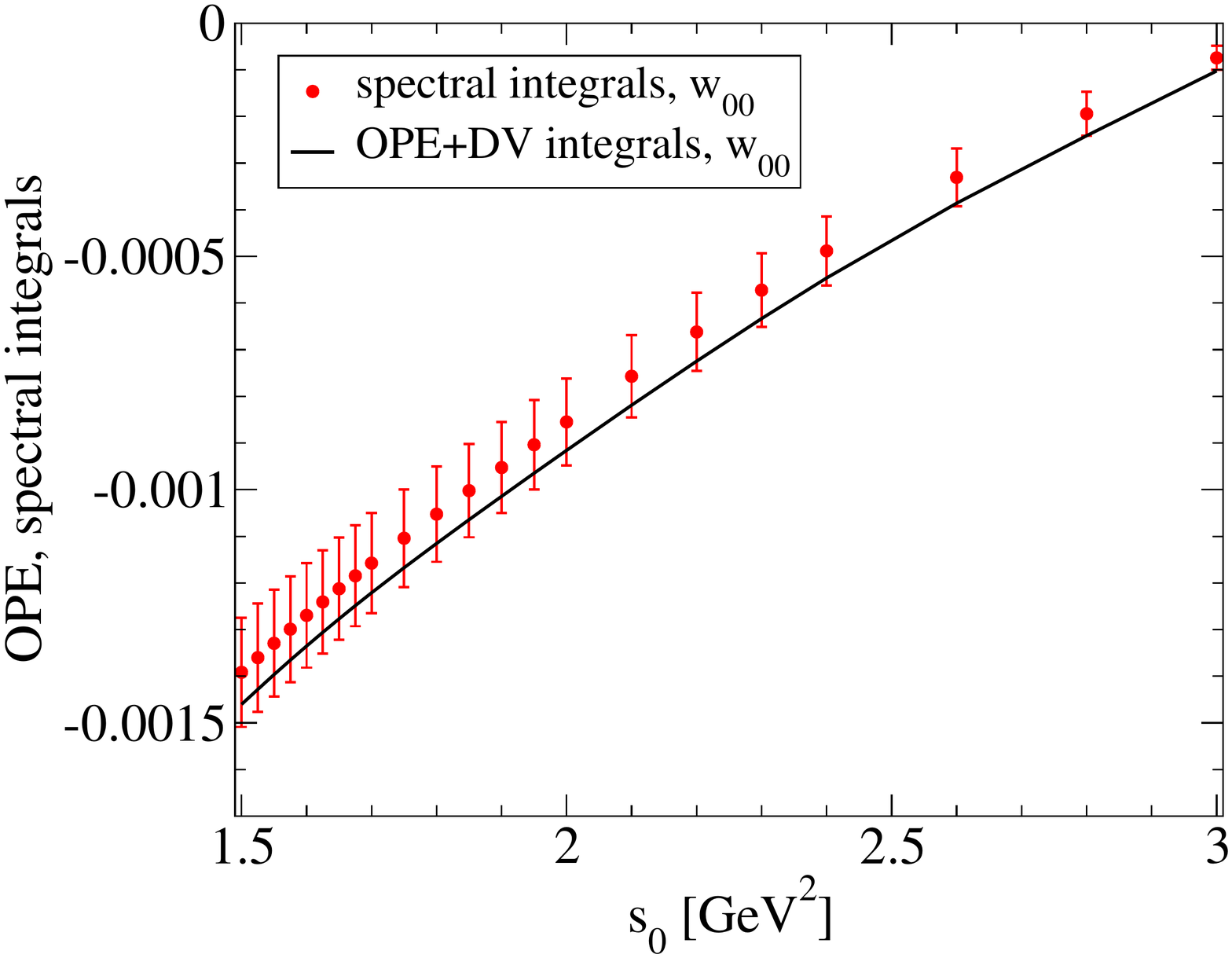}
\hspace{0.0cm}
\includegraphics*[width=7.5cm]{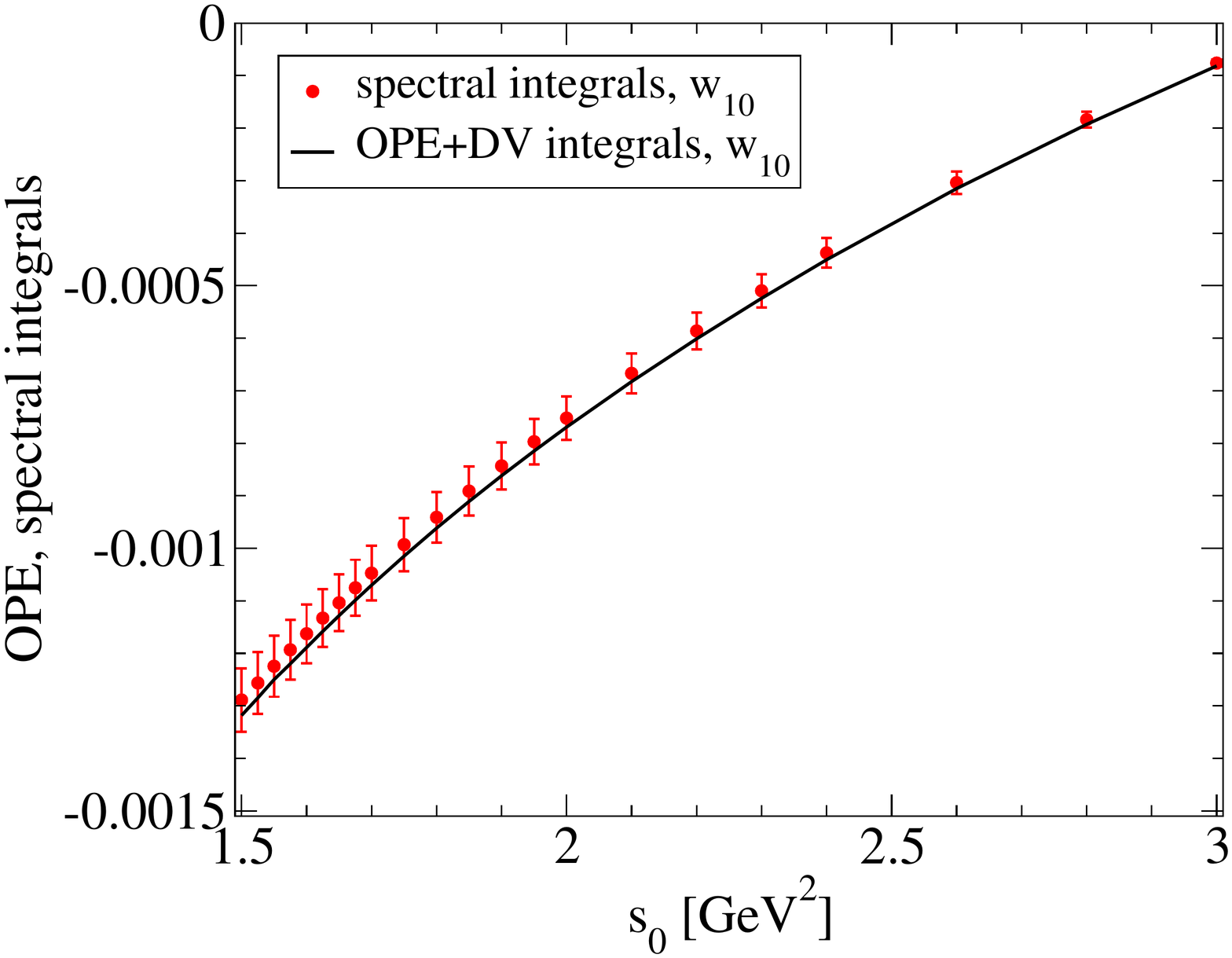}
\vspace{1cm}
\includegraphics*[width=7.5cm]{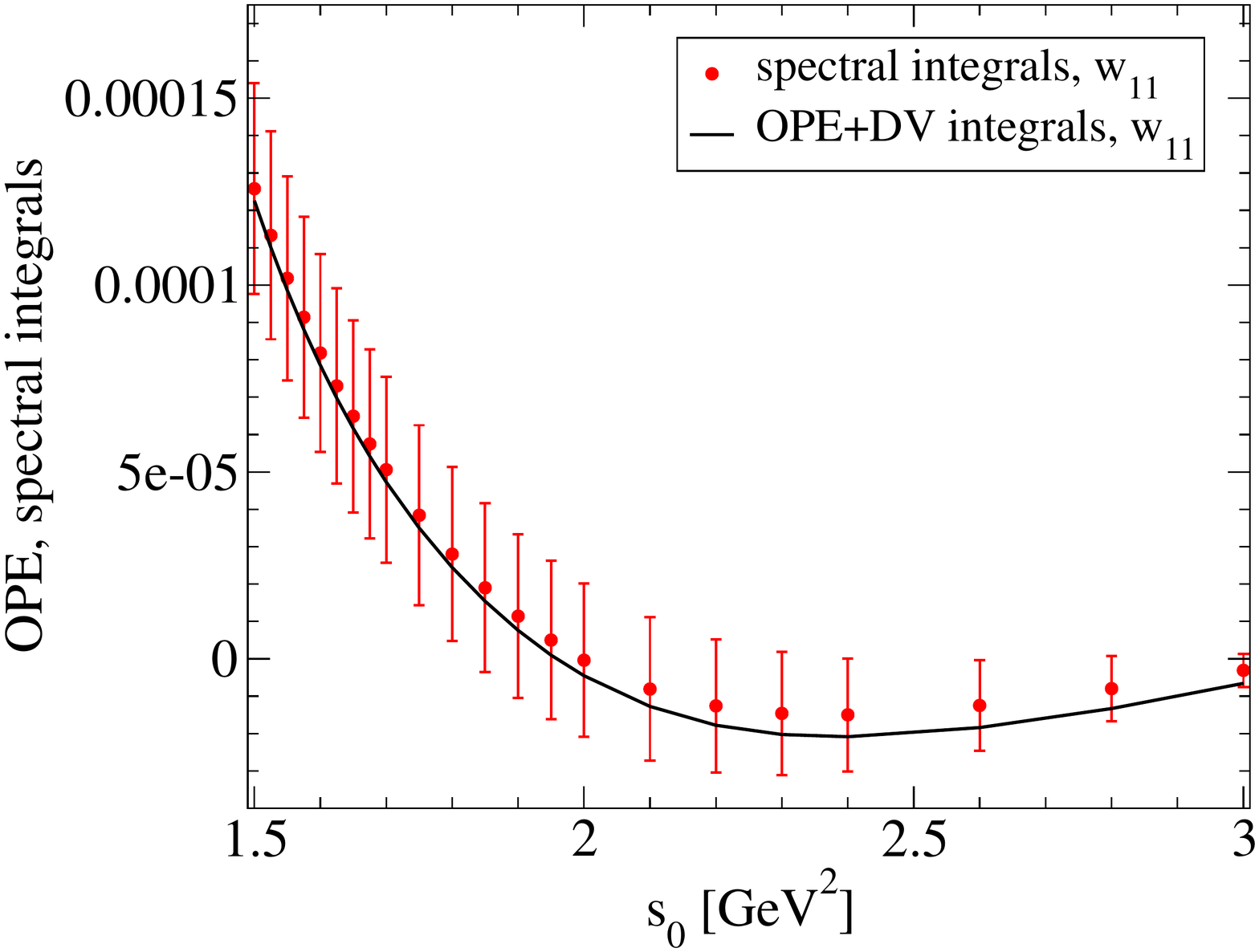}
\hspace{0.0cm}
\includegraphics*[width=7.5cm]{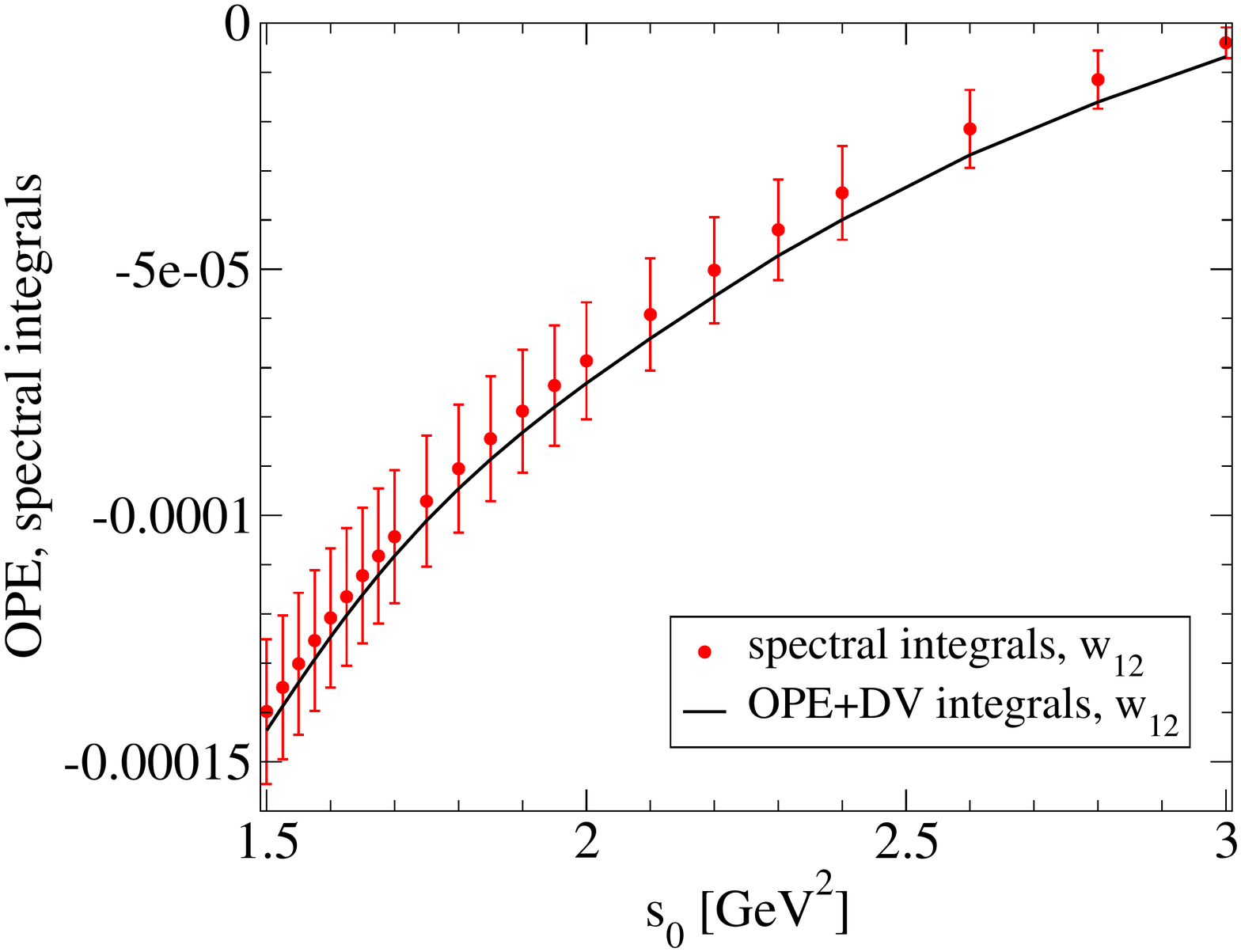}
\begin{minipage}{14pc}
\vspace{-2cm}
\includegraphics*[width=7.5cm]{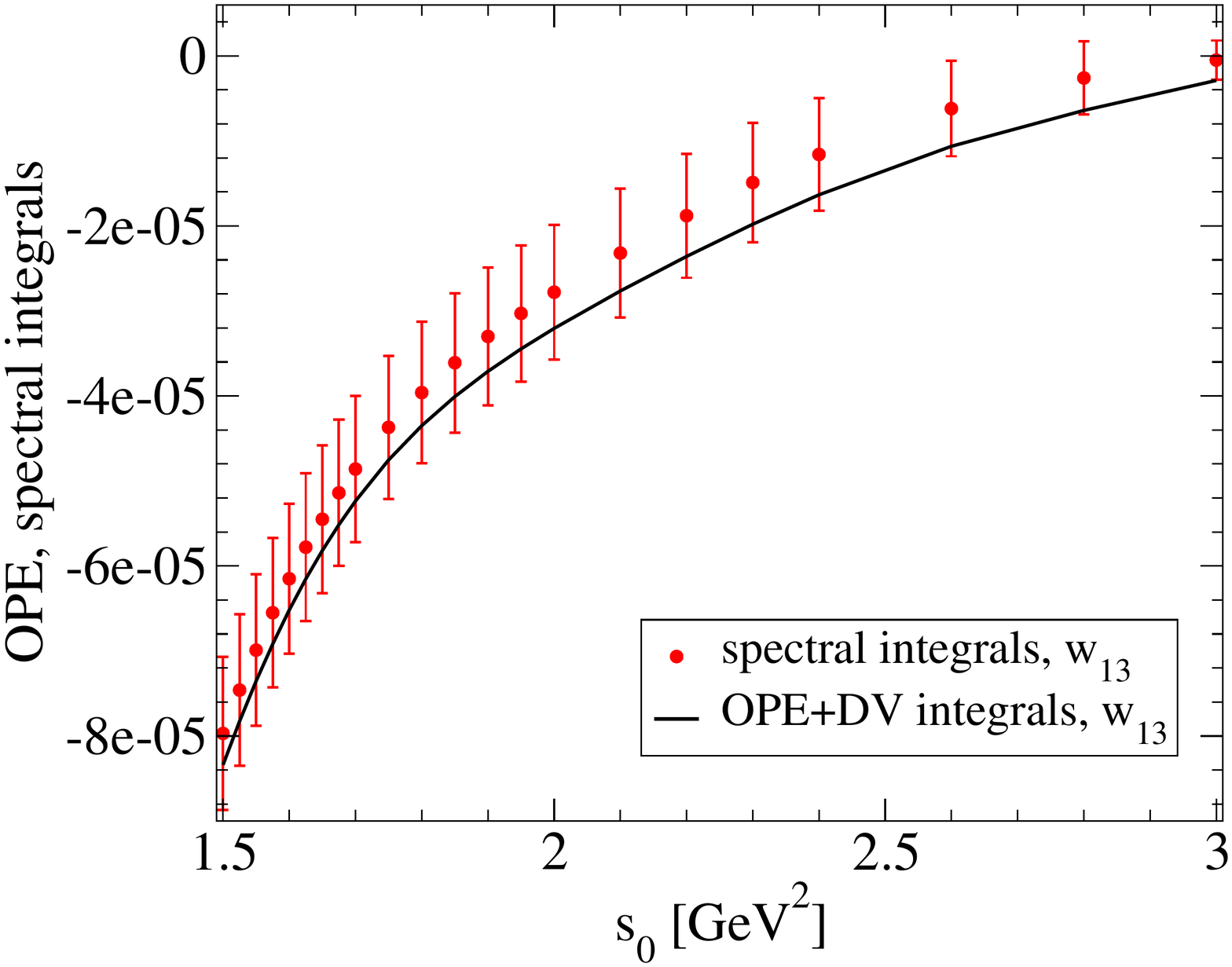}
\end{minipage}\hspace{6pc}
\begin{minipage}[b]{14pc}\caption{\label{ALEPHDVmodel}
{\it Comparison between data and fitted theory for the moments of
Eq.~(\ref{ALEPHweights}).  Fits using the DV-model approach.
Shown are differences between the values at $s_0=m_\tau^2$
minus the values at varying $s_0$.}}
\end{minipage}
\vspace*{-4ex}
\end{figure}

A test that is of particular interest is to compare the moments with weights~(\ref{ALEPHweights}) between the fits, which allow us to determine
all OPE coefficients \cite{boito2015}, and the data.    These comparisons are
shown in Fig.~\ref{ALEPHDVmodel}.  The red points show the data, {\it i.e.},
the left-hand side of Eq.~(\ref{fesr}) (``spectral integrals''), while the black
curves show the fitted theory, {\it i.e.},
the right-hand side of Eq.~(\ref{fesr}) (``OPE$+$DV integrals'').
 What is actually shown is the value
of these moments evaluated at $s_0=m_\tau^2$ minus the moment at varying
$s_0$, in order to account for the strong correlations between both the theory integrals at different $s_0$ and the data integrals at different $s_0$.
We note that there is good agreement for all $s_0$
shown in the figures.

\begin{figure}[t!]
\includegraphics*[width=7.5cm]{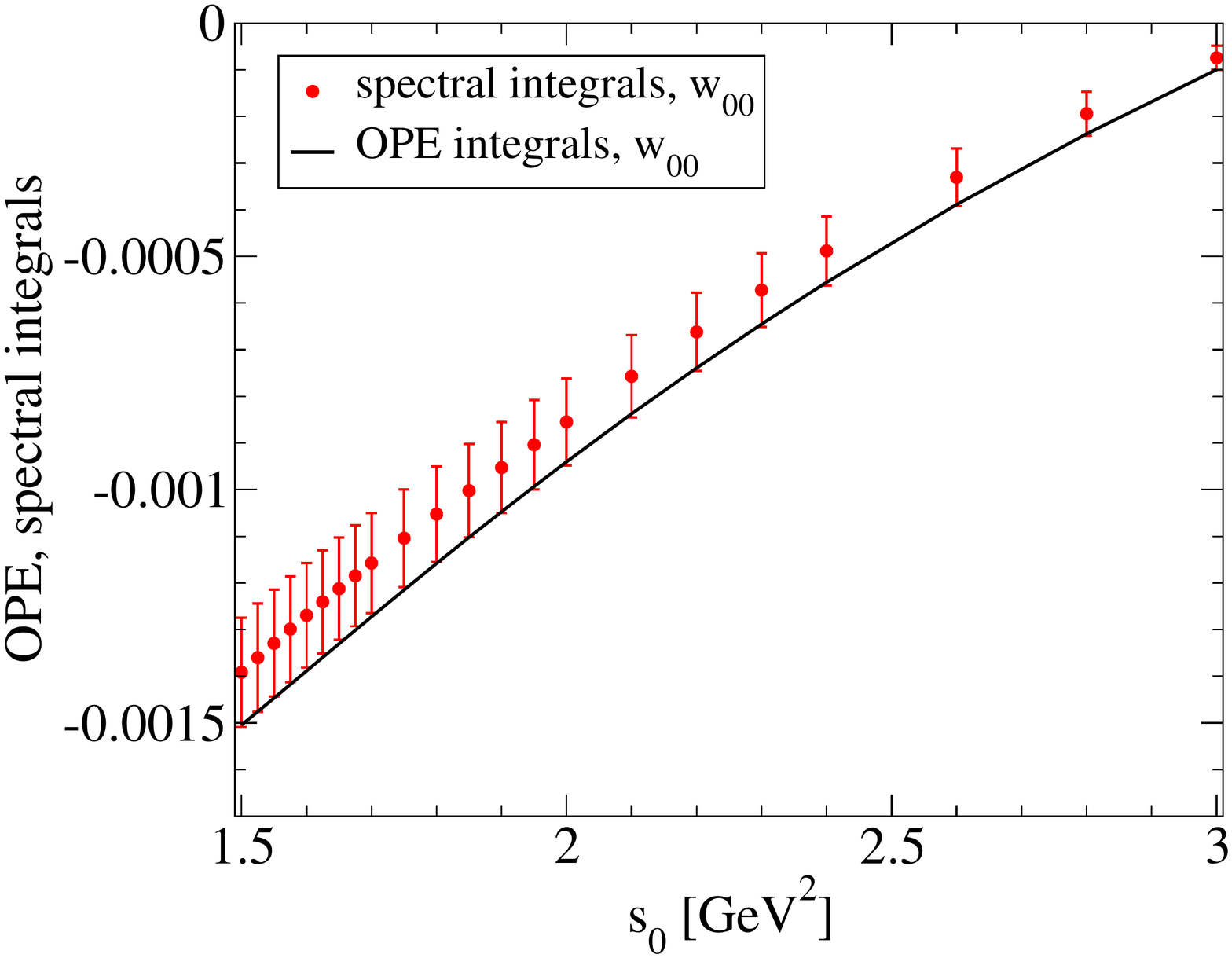}
\hspace{0.0cm}
\includegraphics*[width=7.5cm]{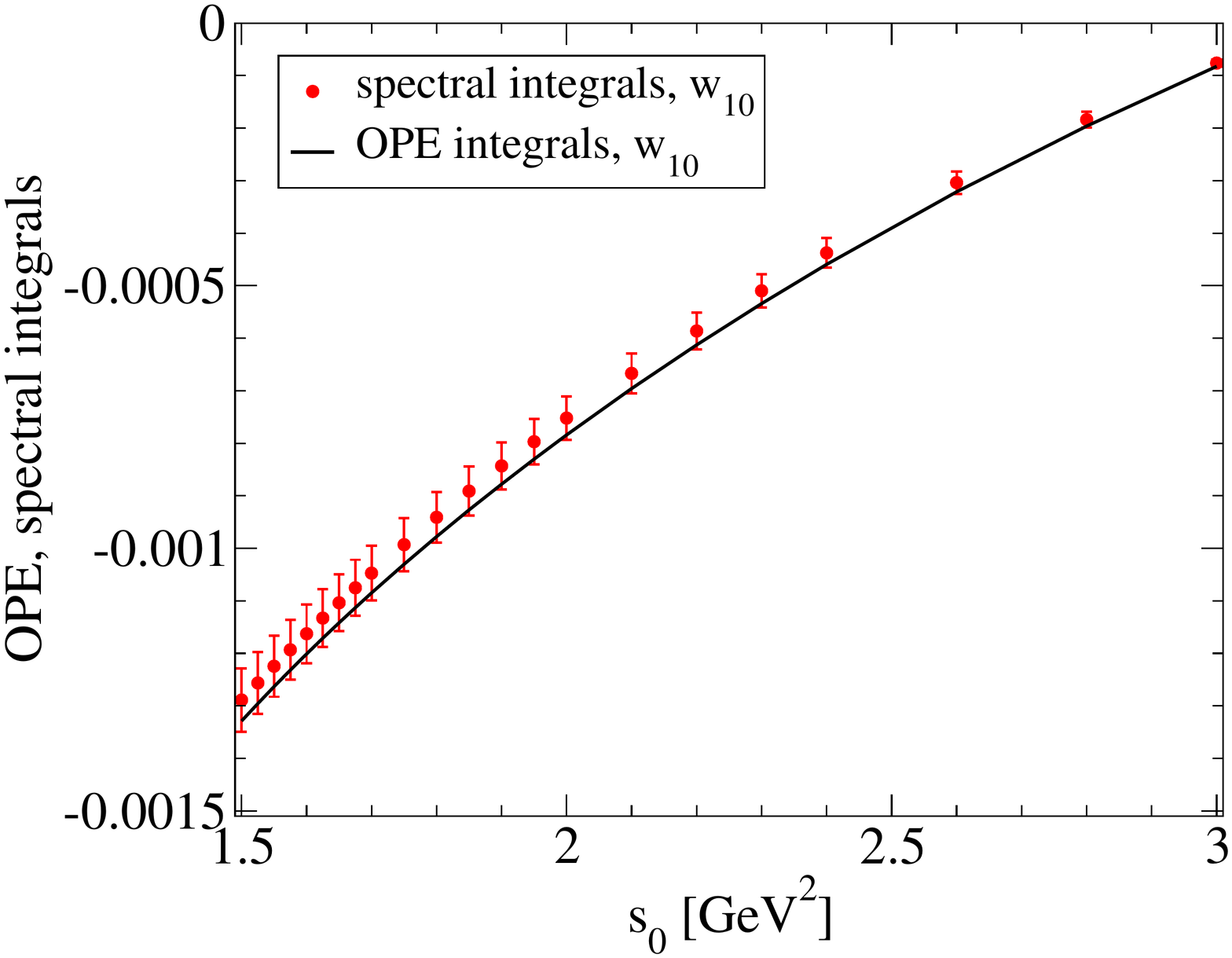}
\vspace{1cm}
\includegraphics*[width=7.5cm]{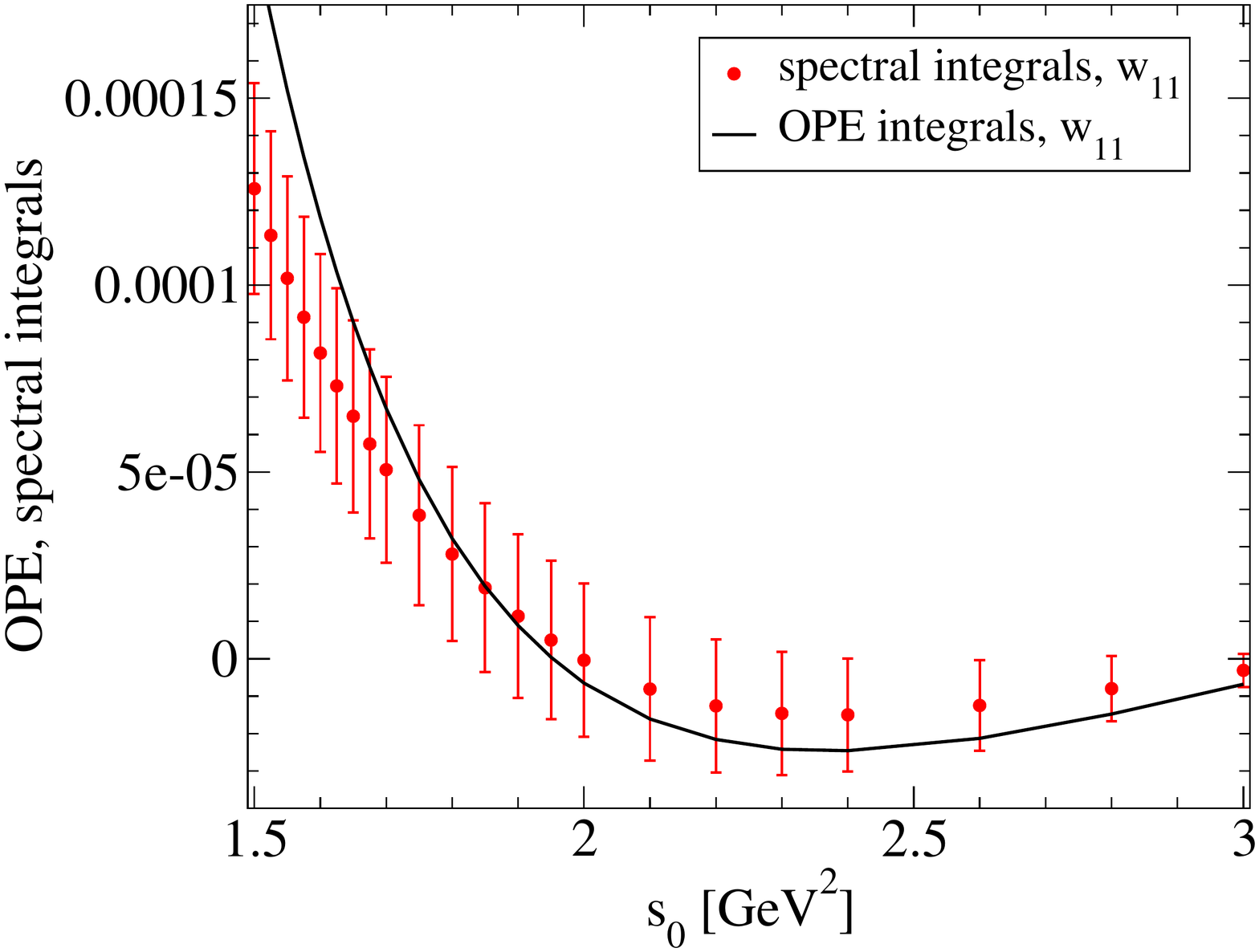}
\hspace{0.0cm}
\includegraphics*[width=7.5cm]{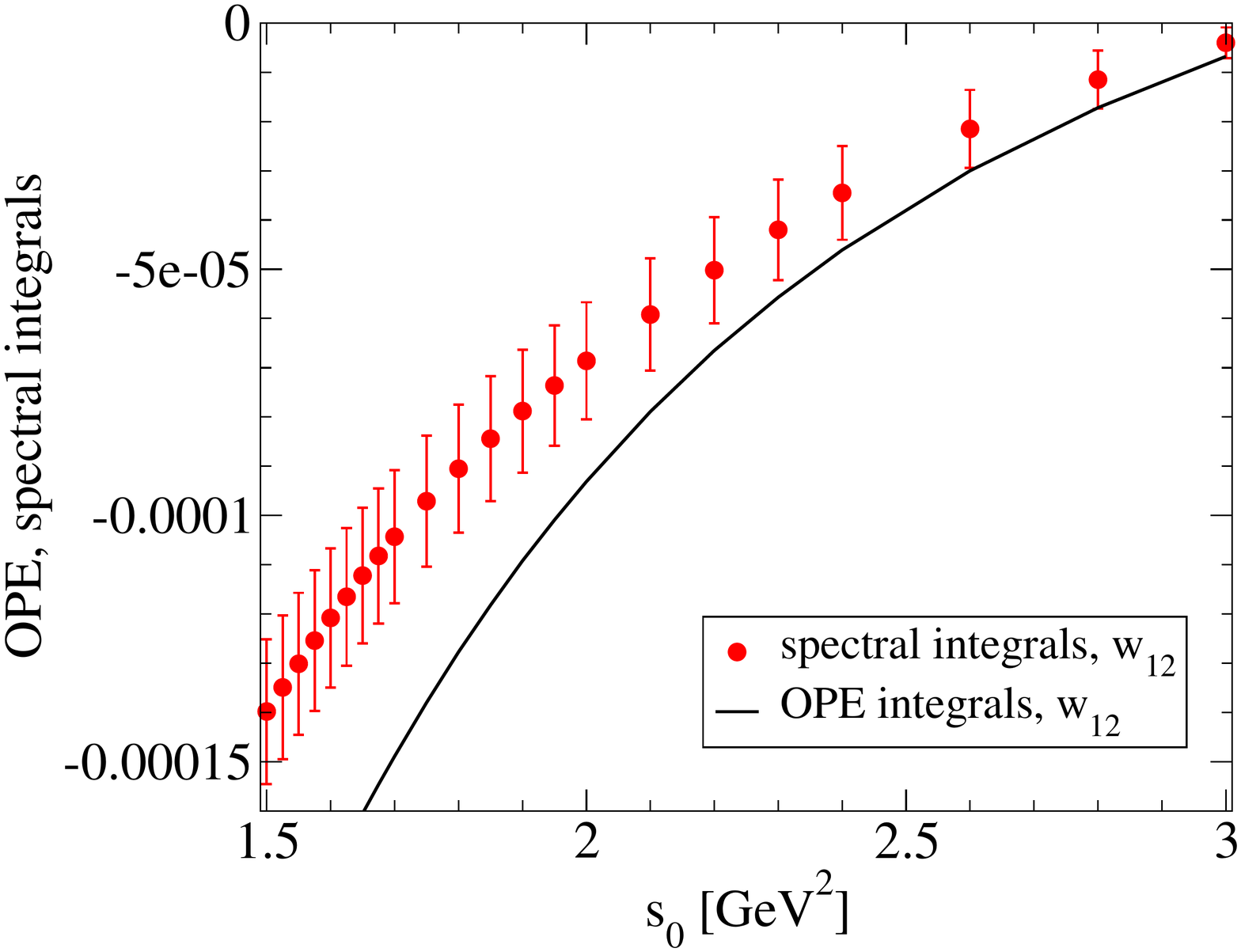}
\begin{minipage}{14pc}
\vspace{-2cm}
\includegraphics*[width=7.5cm]{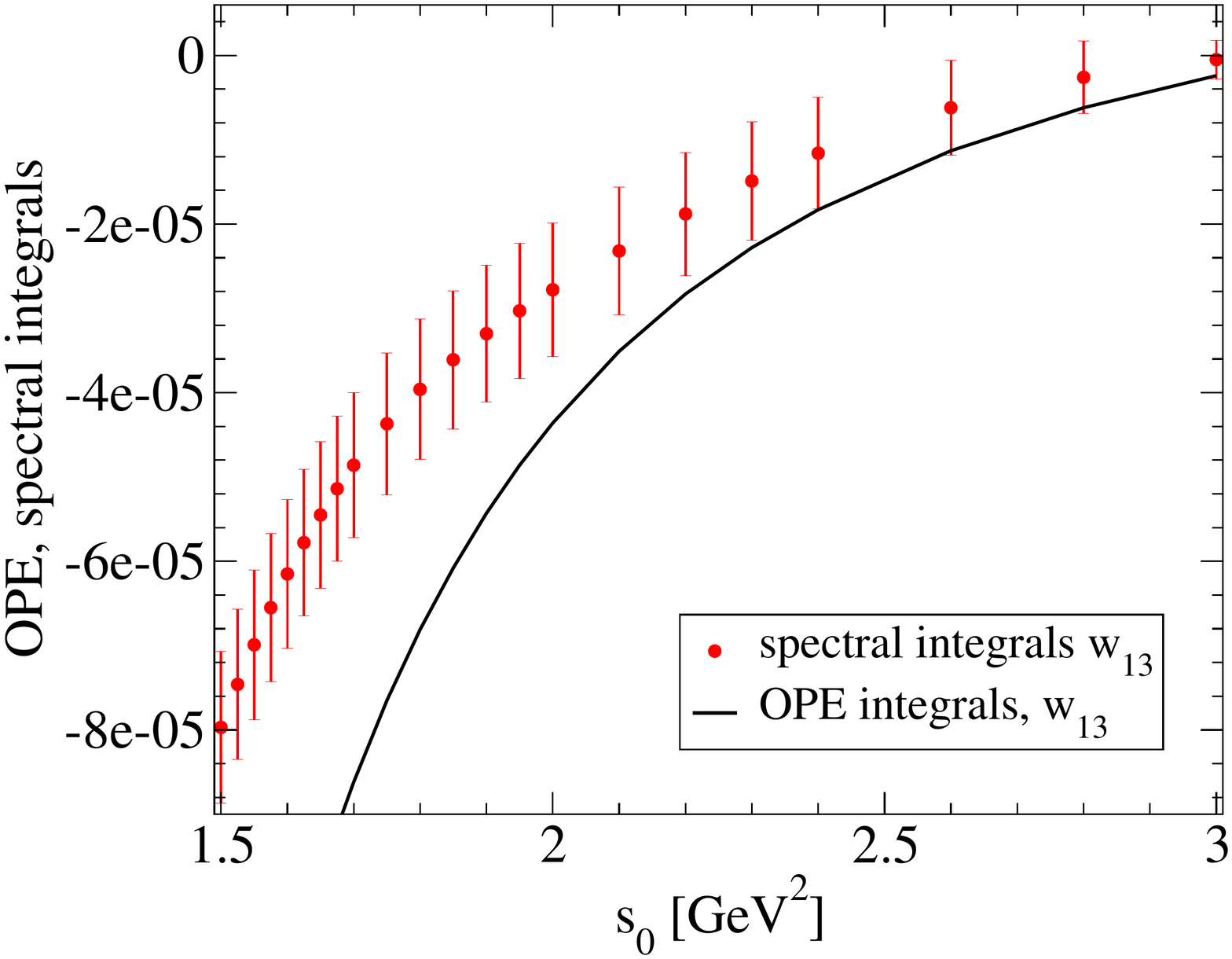}
\end{minipage}\hspace{6pc}
\begin{minipage}[b]{14pc}\caption{\label{ALEPHtruncatedOPE}
{\it Comparison between data and fitted theory for the moments of
Eq.~(\ref{ALEPHweights}).  Fits using the truncated-OPE approach.
Shown are differences between the values at $s_0=m_\tau^2$
minus the values at varying $s_0$.}}
\end{minipage}
\vspace*{-4ex}
\end{figure}

In Fig.~\ref{ALEPHtruncatedOPE} we show the same comparison, but now
using the fit results obtained with the truncated-OPE approach, where all 
moments shown were employed in the fit.   The black curves now only show 
the perturbative plus OPE contribution (``OPE integrals''), because duality
violations, represented by the last term on the right-hand side of Eq.~(\ref{fesr}),
are assumed negligible in the truncated-OPE approach.  The agreement between data and
fitted theory is much less good than for the DV-model approach, for which 
the corresponding comparisons are shown in Fig.~\ref{ALEPHDVmodel}.
Of course, near $s_0=m_\tau^2$ data and fitted theory agree, because of the
differences which are shown.   
Differences between theory and experimental integrals are, of course, by definition, zero at $s_0=m_\tau^2$; 
but deviation of the theory curves from the data  at lower $s_0$ signals shortcomings in the theoretical representation.
  For the moments with weights $w_{12}$
and $w_{13}$, the agreement already deteriorates well above $s_0=2$~GeV$^2$.   For more detail, we refer to Ref.~\cite{boito2016}.

\section{Discussion and outlook}
\medskip
The determination of the strong coupling, $\alpha_s$, from hadronic $\tau$
decays is interesting for at least two reasons.   First, it provides a 
value of $\alpha_s$ at the $Z$ mass which is quite competitive in comparison
with other methods.   The reason for this is that because of the running, relatively modest errors get reduced when the obtained values for $\alpha_s$ are converted to values at the $Z$ mass.   Second, because it is at present the 
only determination with rather high precision at a rather low energy, it provides a nice test of the running of the strong coupling predicted by QCD.   With the presently
available precision, the QCD prediction is nicely confirmed, as indicated, for
instance, by the values shown in Eq.~(\ref{pdgvalues}).

However, as we have discussed in detail, the determination from $\tau$ decays
is afflicted by contamination from non-perturbative effects, because of the rather
low value of the $\tau$ mass.   Some of these non-perturbative effects are captured by the OPE, but not all:   resonance effects are clearly important, and
appear as significant deviations from the quark-gluon picture underlying 
perturbation theory and the OPE.   Another, probably related, issue is the 
discrepancy between values obtained using CIPT or FOPT, which correspond to two different resummation schemes.\footnote{Other resummation schemes are possible, and have been considered.}

Two methods for extracting $\alpha_s$ from $\tau$ decays have been pursued.
One of these, the truncated-OPE approach, has now been shown to fail; it cannot, in general, account for all non-perturbative effects that afflict the analysis.   The
other approach is based on a model for duality violations, and has thus far been shown to be internally consistent, and consistent with the data.   However, it is
important to test this approach more stringently, and much more precise data
for the $V$ channel would provide the possibility to carry out such tests, and,
if successful, lead to a possible reduction in the systematic error.\footnote{For other ideas on possible tests, see Ref.~\cite{caprini2014}.}   With thousands times more $\tau$ pairs, inclusive spectral functions obtained from
Belle and Belle-II data would almost certainly have a significant impact on our understanding of QCD from hadronic $\tau$ decays.   Even a precise determination of the $\tau\to\nu_\tau 4\pi$ channels might already have an 
impact, as the large errors in the large-$s$ region on the currently available $V$ spectral functions are dominated by these channels.   In this respect, we believe that it will suffice to focus on the non-strange $V$ channel, since, contrary to popular belief, there is no indication that a higher
precision can be obtained from $V+A$. It is not unlikely that the 
asymptotic regime where the right-hand side of Eq.~(\ref{fesr}) is valid sets in 
at lower values of $s$ in the $V$ than in the $A$ channel.

It is more difficult to make progress on the optimal resummation scheme.
The typical difference between values of $\alpha_s(m_\tau^2)$ obtained 
from FOPT or CIPT is always about 5\%, with CIPT providing the higher
value, irrespective of the analysis approach employed.   Theoretical
considerations based on a renormalon analysis somewhat favor FOPT
\cite{beneke2008,beneke2013}, but other resummation schemes cannot
be ruled out \cite{caprini2009}.   A direct comparison between the value of $\alpha_s$ 
from $\tau$ decays and from the lattice, for instance, tends to favor 
the CIPT-based value.   Again, higher precision determinations based on 
higher precision data may contribute to a deeper understanding.

\ack
\medskip
MG would like to thank the organizers of the Mini Workshop on Tau Physics
for the opportunity to present this work, and for warm hospitality.
The work of DB is supported by the S{\~a}o Paulo Research Foundation 
(FAPESP) Grant No. 2015/20689-9 and by CNPq Grant No. 305431/2015-3.
 This material is based 
upon work supported by the U.S. Department of Energy, Office of Science, 
Office of High Energy Physics, under Award Number DE-FG03-92ER40711
(MG). 
KM is supported by a grant from the Natural Sciences and Engineering Research
Council of Canada.  SP is supported by CICYTFEDER-FPA2014-55613-P, 2014-SGR-1450.


\section*{References}
\begin{thebibliography}{9}

\bibitem{jamin2005}
  M.~Jamin,
  JHEP {\bf 0509}, 058 (2005)
  [hep-ph/0509001].
  
\bibitem{aleph}
  R.~Barate {\it et al.} [ALEPH Collaboration],
  Eur.\ Phys.\ J.\  C\ {\bf 4}, 409 (1998);
S.~Schael {\it et al.}  [ALEPH Collaboration],
  Phys.\ Rept.\  {\bf 421}, 191 (2005)
  [arXiv:hep-ex/0506072];

\bibitem{davier2014}
  M.~Davier, A.~Hoecker, B.~Malaescu, C.~Z.~Yuan and Z.~Zhang,
  Eur.\ Phys.\ J.\ C {\bf 74}, 2803 (2014)
  [arXiv:1312.1501 [hep-ex]].

\bibitem{pich2016}
  A.~Pich and A.~Rodr{\'i}guez-S{\'a}nchez,
  Phys.\ Rev.\ D {\bf 94}, no. 3, 034027 (2016)
  [arXiv:1605.06830 [hep-ph]].

\bibitem{boito2015}
  D.~Boito, M.~Golterman, K.~Maltman, J.~Osborne and S.~Peris,
  Phys.\ Rev.\ D {\bf 91}, 034003 (2015)
  [arXiv:1410.3528 [hep-ph]].

\bibitem{opal}
K.~Ackerstaff {\it et al.}  [OPAL Collaboration],
  Eur.\ Phys.\ J.\  C\ {\bf 7} (1999) 571
  [arXiv:hep-ex/9808019].

\bibitem{boito2012b}
  D.~Boito, M.~Golterman, M.~Jamin, A.~Mahdavi, K.~Maltman, J.~Osborne and S.~Peris,
  Phys.\ Rev.\ D {\bf 85}, 093015 (2012)
  [arXiv:1203.3146 [hep-ph]].

\bibitem{pdg}
  C.~Patrignani {\it et al.} [Particle Data Group],
  Chin.\ Phys.\ C {\bf 40}, no. 10, 100001 (2016).

\bibitem{baikov2008}
P.~A.~Baikov, K.~G.~Chetyrkin and J.~H.~K\"uhn,
  Phys.\ Rev.\ Lett.\  {\bf 101} (2008) 012002
  [arXiv:0801.1821 [hep-ph]].
  
\bibitem{braaten1992}
E.~Braaten, S.~Narison, and A.~Pich,
 Nucl.\ Phys.\ B\ {\bf 373} (1992) 581.
 
\bibitem{tsai1971}
  Y.-S. Tsai, Phys.\ Rev.\ D\ {\bf 4}, 2821 (1971).
  
\bibitem{boito2012}
  D.~Boito, O.~Cat\`a, M.~Golterman, M.~Jamin, K.~Maltman, J.~Osborne and S.~Peris,
  Phys.\ Rev.\ D\ {\bf 84}, 113006 (2011)
  [arXiv:1110.1127 [hep-ph]].

\bibitem{cata2005}
  O.~Cat\`a, M.~Golterman, S.~Peris,
  JHEP {\bf 0508}, 076 (2005)
  [hep-ph/0506004].

\bibitem{cata2008}
  O.~Cat\`a, M.~Golterman, S.~Peris,
  Phys.\ Rev.\  D\ {\bf 77}, 093006 (2008)
  [arXiv:0803.0246 [hep-ph]].
  
\bibitem{diberder1992}
F.~Le Diberder, A.~Pich,
  Phys.\ Lett.\  B\ {\bf 289}, 165 (1992).
  
\bibitem{boito2016}
  D.~Boito, M.~Golterman, K.~Maltman and S.~Peris,
  Phys.\ Rev.\ D {\bf 95}, no. 3, 034024 (2017)
  [arXiv:1611.03457 [hep-ph]].
  
\bibitem{maltman2008}
  K.~Maltman and T.~Yavin,
  Phys.\ Rev.\ D {\bf 78}, 094020 (2008)
  doi:10.1103/PhysRevD.78.094020
  [arXiv:0807.0650 [hep-ph]].

\bibitem{cata2009}
  O.~Cat\`a, M.~Golterman and S.~Peris,
  Phys.\ Rev.\ D {\bf 79}, 053002 (2009)
  doi:10.1103/PhysRevD.79.053002
  [arXiv:0812.2285 [hep-ph]].

\bibitem{blok1998}
B.~Blok, M.~A.~Shifman and D.~X.~Zhang,
  Phys.\ Rev.\  D\ {\bf 57}, 2691 (1998)
  [Erratum-ibid.\  D\ {\bf 59}, 019901 (1999)]
  [arXiv:hep-ph/9709333].

\bibitem{bigi1999}
  I.~I.~Y.~Bigi, M.~A.~Shifman, N.~Uraltsev, A.~I.~Vainshtein,
  Phys.\ Rev.\  D\ {\bf 59}, 054011 (1999)
  [hep-ph/9805241].
  
\bibitem{golterman2002}
  M.~Golterman, S.~Peris, B.~Phily, E.~de Rafael,
  JHEP {\bf 0201}, 024 (2002)
  [hep-ph/0112042].

\bibitem{caprini2014}
  I.~Caprini, M.~Golterman and S.~Peris,
  Phys.\ Rev.\ D\ {\bf 90}, 033008 (2014)
  [arXiv:1407.2577 [hep-ph]];
    D.~Boito and I.~Caprini,
  Phys.\ Rev.\ D {\bf 95}, no. 7, 074027 (2017)
  doi:10.1103/PhysRevD.95.074027
  [arXiv:1702.03757 [hep-ph]].

\bibitem{beneke2008}
M.~Beneke and M.~Jamin,
  JHEP {\bf 0809}, 044  (2008)  [arXiv:0806.3156 [hep-ph]].
  
\bibitem{beneke2013}
  M.~Beneke, D.~Boito and M.~Jamin,
  JHEP {\bf 1301}, 125 (2013)
  [arXiv:1210.8038 [hep-ph]].

\bibitem{caprini2009}
See, for instance,  I.~Caprini and J.~Fischer,
  Eur.\ Phys.\ J.\  C\ {\bf 64}, 35 (2009)
  [arXiv:0906.5211 [hep-ph]];
    G.~Abbas, B.~Ananthanarayan, I.~Caprini and J.~Fischer,
  Phys.\ Rev.\ D\ {\bf 87}, 014008 (2013)
  [arXiv:1211.4316 [hep-ph]];
    I.~Caprini,
  Mod.\ Phys.\ Lett.\ A\ {\bf 28}, 1360003 (2013)
  [arXiv:1306.0985 [hep-ph]];
  G.~Abbas, B.~Ananthanarayan and I.~Caprini,
  Mod.\ Phys.\ Lett.\ A\ {\bf 28}, 1360004 (2013)
  [arXiv:1306.1095 [hep-ph]].

\end{thebibliography}
\end{document}